\documentclass{article}
\usepackage{amssymb,amsfonts}
\usepackage[intlimits]{amsmath}
\usepackage{verbatim}
\usepackage{mathbbol}
\usepackage{bbm}
\usepackage{mathrsfs}

\newcommand{\<}{\langle}
\renewcommand{\>}{\rangle}
\newcommand{\wh}[1]{\widehat{#1}}

\newcommand{\bra}[1]{\left\langle {#1} \right|}
\newcommand{\ket}[1]{ \left| {#1} \right\rangle}
\newcommand{\braket}[2]{\left\langle {#1} \left| \right. {#2} \right\>}
\newcommand{\la}{\lambda}

\newcommand{\ve}{\varepsilon}
\newcommand{\vp}{\varphi}
\newcommand{\si}{\sigma}
\newcommand{\ep}{\epsilon}
\newcommand{\con}{\textrm{const}}
\newcommand{\lime}{\lim_{\epsilon \rightarrow 0^+}}
\newcommand{\liml}{\lim_{\la \rightarrow 0^+}}
\newcommand{\mc}[1]{\mathcal{#1}}
\newcommand{\mbb}[1]{\mathbbm{#1}}
\newcommand{\mr}[1]{\mathrm{#1}}

\newcommand{\tr}[1]{\textrm{#1}}

\newcommand{\ol}[1]{\overline{#1}}
\newcommand{\q}{\quad}

\DeclareMathOperator{\Realis}{\mathfrak{Re}}
\DeclareMathOperator{\Imaginalis}{\mathfrak{Im}}
\newcommand{\Tr}{\textrm{Tr}}

\title{Global vs local Casimir effect}
 \author{Andrzej Herdegen\thanks{e-mail: herdegen@th.if.uj.edu.pl}
 \quad and\quad Mariusz Stopa\\
{\it Institute of Physics, Jagiellonian University,}\\
{\it Reymonta 4, 30-059 Cracow, Poland}}
\date{}

\begin{document}

\maketitle

\begin{abstract}
This paper continues the investigation of the Casimir effect
with the use of the algebraic formulation of quantum field
theory in the initial value setting.

Basing on earlier papers by one of us (AH) we approximate the
Dirichlet and Neumann boundary conditions by simple interaction
models whose nonlocality in physical space is under strict
control, but which at the same time are admissible from the
point of view of algebraic restrictions imposed on models in
the context of Casimir backreaction. The geometrical setting is
that of the original parallel plates. By scaling our models and
taking appropriate limit we approach the sharp boundary
conditions in the limit. The global force is analyzed in that
limit. One finds in Neumann case that although the sharp
boundary interaction is recovered in the norm resolvent sense
for each model considered, the total force per area depends
substantially on its choice and diverges in the sharp boundary
conditions limit. On the other hand the \emph{local} energy
density outside the interaction region, which in the limit
includes any compact set \emph{outside the strict position of
the plates}, has a universal limit corresponding to sharp
conditions. This is what one should expect in general, and the
lack of this discrepancy in Dirichlet case is rather
accidental. Our discussion pins down its precise origin: the
difference in the order in which scaling limit and integration
over the whole space is carried out.

\end{abstract}
\vfill \eject

\sloppy

\section{Introduction and the main idea}\label{intII}

The most natural setting for the consideration of the Casimir
effect is the algebraic approach. This approach allows a
mathematically rigorous analysis of the effect and gives a
clear understanding of the sources of the difficulties one
encounters in more traditional treatments. In application to
quantum fields this analysis rests, in broad terms, on the
following cornerstones.\\
 (i) A quantum relativistic theory is defined by
an algebra of observables, in simple cases defined directly by
`fields' (scalar, electromagnetic).\\
 (ii) Each particular physical system obeying this
theory is described by a Hilbert space representation of this
algebra. Inequivalent representations refer to physically
non-comparable systems or idealizations (such as a local
isolated system and a~thermodynamic limit system).\\
 (iii) The change of external conditions under which a quantum
system is placed leads to a change of the state of the system.

For the calculation of the global Casimir-type effects, as the
backreaction force, one needs models which respect the above
three constituents of a quantum theory. Thus the model of a
quantum field should be based on one definite algebra, and the
interaction with the external conditions should not lead to a
change of its representation. If this condition is fulfilled
then the Casimir force results from the change in the
expectation value of \emph{one and the same energy observable,
as defined by free field}, as the state changes with changing
external conditions (such as position of macroscopic bodies).

This analysis has been conducted at length by one of us in
\cite{her05} and \cite{her06}, where also clear cut criterions
for the admissibility of external interaction models for
a~class of systems were formulated. In application to the free
quantum scalar field these amount to the following. Let
$\phi(x)$ be a scalar field and denote by $h^2$ the standard
selfadjoint extension in $L^2(\mbb{R}^3,d^3x)$ of $-\Delta$.
The free field dynamics is then
\begin{equation}\label{free}
 (\partial_t^2+h^2)\phi(t,\vec{x})=0\,.
\end{equation}
Suppose now that one introduces to the system external
macroscopic bodies, such as conducting plates in the original
Casimir system, which change the dynamics of the field. Let $a$
denote free parameters of these bodies, such as separation of
the plates, and let the modified dynamics (for fixed $a$) be
given by
\begin{equation}\label{modified}
 (\partial_t^2+h_a^2)\phi(t,\vec{x})=0\,,
\end{equation}
where $h_a^2$ is a positive selfadjoint operator. Now, the two
settings can be described by one choice of observables algebra,
and in common representation of this algebra if, and only if,
$h_a^{-1/2}(h_a-h)h^{-1/2}$ is a Hilbert-Schmidt operator in
$L^2(\mbb{R}^3,d^3x)$, that is
\begin{equation}\label{numbera}
 \Tr\,\Big[h^{-1/2}(h_a-h)h_a^{-1}(h_a-h)h^{-1/2}\Big]<\infty\,.
\end{equation}
Suppose this condition holds and let the algebra of the field
be represented in some Hilbert space $\mc{H}$. Let further $H$
be the energy operator \emph{as defined by free field
dynamics}, and let $\Omega_a$ be the minimal energy state
vector \emph{as defined by the modified dynamics}
(\ref{modified}). The Casimir energy is then given by
\begin{equation}\label{energya}
 \mc{E}_a=(\Omega_a, H\Omega_a)=
 \frac{1}{4}\Tr\,\Big[(h_a-h)h_a^{-1}(h_a-h)\Big]\,.
\end{equation}
That this energy be finite is another condition on the model of
$h_a$, and only if both conditions are satisfied the Casimir
problem has a finite solution and the Casimir force is then
\begin{equation*}
 \mc{F}_a=-\frac{d\mc{E}_a}{da}\,.
\end{equation*}

These admissibility conditions say, roughly, that the modified
dynamics $h_a^2$ cannot differ much from the free dynamics
$h^2$. Introducing sharp boundary conditions, such as
Dirichlet/Neumann conditions on plates, violates these demands.
One faces therefore the problem of an appropriate approximation
for the description of such plates.

We consider the simplest geometrical situation, the original
Casimir problem of two infinite, parallel plane plates at a
distance $a$ from each other. We assume that $z$-axis is
perpendicular to the planes and the modification of dynamics
affects this direction only, thus
\begin{equation*}
 h^2=h_z^2+h_\perp^2\,,\quad
 h_a^2=h_{za}^2+h_\perp^2\,.
\end{equation*}
where $h_\perp^2$ is the free dynamics in the directions
perpendicular to $z$-axis and $h_{za}^2$ is a modification of
$h_z^2=-\partial_z^2$ in $L^2(\mbb{R},dz)$; we refer the reader
for details to \cite{her06}. In this setting the conditions of
finiteness of (\ref{numbera}) and (\ref{energya}) cannot be
expected to hold as they stand because of translation symmetry
in the planes, and must be replaced by conditions `per unit
area' of the planes. It has been shown in \cite{her06} that
this amounts to
\begin{gather}
 \Tr\,\Big[(h_{za}-h_z)^2\Big]<\infty\,,\label{HS1}\\
 \Tr\,\Big[(h_{za}-h_z)h_z(h_{za}-h_z)\Big]<\infty\,,\label{HS2}
\end{gather}
where the trace refers to the Hilbert space $L^2(\mbb{R},dz)$.
If these conditions are satisfied, the energy per unit area is
finite and reads
\begin{equation}\label{main_energy}
 \ve_a=\frac{1}{24\pi}
 \Tr\,\Big[(h_{za}-h_z)(2h_z+h_{za})(h_{za}-h_z)\Big]\,.
\end{equation}

A class of models for $h_{za}^2$ imitating the boundary
conditions, but consistent with the above demands was
considered in \cite{her06}. The idea was to take
$h_{za}^2=(function\ of\/)(h_z,h_{za}^B)$, where $(h_{za}^B)^2$
is $-\partial_z^2$ with boundary conditions at $z=\pm a/2$. The
choice of functions assured that for small spectral values of
$h_z^2$ and $(h_{za}^B)^2$ the models reproduced the sharp
boundaries, while for large spectral values tended to free
dynamics. Moreover, one could introduce a scaling parameter
$\mu$ such that for $\mu\to\infty$ the models approached the
sharp boundaries in the whole spectrum. For the Casimir energy
per area in the rescaled models $\ve^\la_a$ (we prefer to work
with $\la=1/\mu$ here) one then found
\begin{equation}\label{enlim}
 \ve^\la_a=\frac{\ve_\infty}{\la^3}
 +\frac{c}{\la a^2}-\frac{\pi^2}{1440 a^3}+
 (\text{terms}\to0\text{ for }\la\to0)\,,
\end{equation}
where $\ve_\infty$ and $c$ are constants, $c=0$ in Dirichlet
case, but $c\neq0$ in Neumann case. It was also shown that the
direct sum of the two models describes the setting of the
electromagnetic field between conducting plates. Thus the
Casimir force per unit area is then
\begin{equation*}
 -\,\frac{d\,[\ve_a^\la(D)+\ve_a^\la(N)]}{da}
 =\frac{2c}{\la a^3}-\frac{\pi^2}{240a^4}
 +(\text{terms}\to0\text{ for }\la\to0)\,.
\end{equation*}
The second term reproduces the well-known Casimir's formula,
but the first term is model-dependent and dominates for large
$a$. Moreover, in typical situations there is $c>0$ and the
force becomes repulsive for large $a$.

In the present paper we want to find out whether these results
will be confirmed in another class of models, constructed in a
wholly different way. Rather than manipulate spectral
properties directly, now we want to approximate interaction
with the plates directly in the physical space. It is easy to
see that strictly local potential interaction of the form
$[h_{za}^2\psi](z)=-\partial_z^2\psi(z)+V(z)\psi(z)$ violates
our conditions. Therefore we replace $V$ by a slightly nonlocal
integral quasi-potential $[V\psi](z)=\int V(z,z')\psi(z')dz'$
with the kernel $V(z,z')$ concentrated around the position of
the plates. We show that with an appropriately defined scaling
of some simple kernels of this kind one can reproduce sharp
boundary conditions on the plates in the limit. The Casimir
energy can again be calculated and for a class of models the
result (\ref{enlim}) is confirmed. However, in general the
Neumann case proves to be even more singular here than in the
models considered in \cite{her06} and the universal term could
be disturbed. Nonlocal quasi-potentials has been considered in
the Casimir context by other authors before, but in different
formalisms and with rather different motivations (see e.g.\
\cite{nonloc}).

The present choice of models makes also possible a local
analysis of the local energy \emph{density}. We show that
\emph{outside the interaction region} the density tends in the
scaling limit to a well defined universal form corresponding to
sharp boundary conditions. The present mathematically rigorous
setting allows the comparison and better understanding of the
local -- global relation. The model-dependent and divergent (in
the limit) contributions to the global force are due to the
interaction region. We discuss this point more fully in the
Discussion section.

For a more extensive discussion of the background of the
present paper, as well as for more extensive literature we
refer the reader to \cite{her05} and \cite{her06}. We define
our models in Section \ref{models}. Appropriate scaling of
these models is shown to reproduce the sharp boundary
conditions in Section \ref{odtw}. Spectral properties of the
models are discussed in Section \ref{spectral} and the
admissibility of the models in the sense mentioned above is
proved in Section \ref{HS}. It is shown in Section
\ref{scaling} that the Casimir energy of the scaled models is
obtained by the expansion of the formula for energy in inverse
powers of $a$, and this expansion (up to a significant order)
is obtained in Section~\ref{enexp}. For comparison, in Section
\ref{local} we obtain local results and their scaling limit.
The discussion occupies Section \ref{discussion}. More
technical points of our derivations are shifted to Appendices.

\section{The models}\label{models}

We postulate for our analysis the following quasi-potentials
\begin{equation}\label{qp}
 V=\si \Bigl( \ket{U_b g}\bra{U_b g}+\ket{U_{-b}g}
 \bra{U_{-b}g}\Bigr)\,, \quad b=a/2>0 \,,\q
 \si =\pm 1 \,,
\end{equation}
where $U$ is the translation operator and $\si=1,-1$
corresponds to Dirichlet (D) and Neumann (N) conditions
respectively. These conditions will be achieved in the two
cases by an appropriate scaling limit to be defined below. In
all what follows one should keep in mind that unless stated
otherwise we treat parallelly both cases, but the dependence of
quantities on $\si$ is suppressed.

In position representation the quasi-potential is an integral
operator
 $(V\psi)(z) = \int V(z,z')\psi(z')\,dz'$ with the kernel
\begin{equation*}
 V(z,z') =\si\Big[g(z-b)\ol{g(z'-b)} + g(z+b)\ol{g(z'+b)}\Big]\,.
\end{equation*}
For the functions $g$ we assume that
\begin{equation}\label{fDN}
 g(z) =
 \begin{cases}
 f(z) & \text{if $\si=1$},\quad\qquad \text{(D)}\\
 -i \frac{d}{dz} f(z) & \text{if $\si=-1$},\qquad\ \text{(N)}
 \end{cases}
\end{equation}
where $f$ is a complex, compactly supported smooth function,
with the following properties
\begin{gather}
 f(-z)=f(z)\,,\quad \mathrm{supp}f\subseteq\<-R,R\>\,,\ R<b\,,
 \quad \wh{f}(0)\neq0\,,\label{assum}\\[.7ex]
 \|f\|=1\,,\qquad\qquad \text{(N)}\label{normN}
\end{gather}
where $\wh{f}$ is the Fourier-transformed function
\begin{equation*}
 \wh{f}(p)=\frac{1}{\sqrt{2\pi}}\int f(z)e^{-ipz}dz\,,
\end{equation*}
and the last property is assumed only in the Neumann case. The
first condition reflects the symmetry of each of the plates,
the second says that the nonlocalities of the two interaction
centers at $z=\pm b$ do not overlap, and the third and fourth
are technical.

We denote by $h_z$ and $h_{za}$ the selfadjoint, nonnegative
square roots of the operators
\begin{equation}\label{hqp}
 h_z^2=-(d/dz)^2\,,\quad h_{za}^2 = h_z^2 + V\,,
\end{equation}
respectively. Operator $h_z^2$ is the standard one-dimensional
Laplace operator (with opposite sign), while $h_{za}^2$ is its
Kato-Rellich perturbation, with unchanged domain, as $V$ is
bounded. The (strict) positivity of $h_{za}^2$ in the Dirichlet
case is obvious, while in the Neumann case one has by
(\ref{fDN}) that for all $\psi$ in the domain of $h_z^2$ there
is
\begin{equation*}
 (\psi,h_{za}^2\psi)
 =\|\psi'\|^2-|(U_{+b}f,\psi')|^2-|(U_{-b}f,\psi')|^2\,,
 \qquad\qquad\text{(N)}
\end{equation*}
where $\psi'(z)=d\psi(z)/dz$. The functions $U_{\pm b}f$ do not
overlap, and satisfy \mbox{$\|U_{\pm b}f\|=\|f\|$}. Thus by
(\ref{normN}) they form an orthonormal system, which implies
$(\psi,h_{za}^2\psi)\geq0$ for each $\psi$. (Here a more
strongly bounding condition $\|f\|<1$ would produce a strictly
positive operator; this, however, would not lead to the
recovery of Neumann condition in a limit to be defined below.)

For $w^2\in\mbb{C}$ with $\Imaginalis w^2\neq0$ the resolvents
denoted by
\begin{equation}\label{resol}
 G_0(w^2)=(w^2-h_z^2)^{-1}\,,\
 G(w^2)=(w^2-h_{za}^2)^{-1}\,,\
\end{equation}
are bounded operators. In all what follows for given $w^2$ we
fix $w$ by $\Imaginalis w>0$. We also introduce the
$T$-operator known from the stationary scattering theory
\begin{equation}\label{Top}
 T(w^2)=V+VG_0(w^2)T(w^2)\,.
\end{equation}
This equation may be explicitly solved for $T$: making the
Ansatz
\begin{equation}\label{Tal}
 T(w^2)=
 \begin{pmatrix}\,\ket{U_{+b}g}\!&\!\ket{U_{-b}g}\,\end{pmatrix}
 \mc{T}(w^2)
 \begin{pmatrix}\bra{U_{+b}g}\\[1ex] \bra{U_{-b}g}\end{pmatrix}\,,
\end{equation}
with $\mc{T}(w^2)$ a numerical matrix, one easily finds
\begin{equation}\label{Tw2}
 \mc{T}(w^2) =
 \begin{pmatrix}
 \si-(g,G_0(w^2)g)& -(U_ag,G_0(w^2)g)\\[1.5ex]
 -(U_ag,G_0(w^2)g)& \si-(g,G_0(w^2)g)
 \end{pmatrix}^{-1}\,.
\end{equation}
The resolvent may be then expressed by
\begin{equation}\label{GTG}
 G(w^2)=G_0(w^2)+G_0(w^2)\,T(w^2)G_0(w^2)\,.
\end{equation}
In momentum representation, taking into account \eqref{fDN} and
denoting
\begin{equation}\label{FM}
 \mc{F}_p=p^{\frac{1-\si}{2}}\wh{f}(p)
 \begin{pmatrix}e^{-ibp}&e^{+ibp}\end{pmatrix}\,,\quad
 M_p=|\wh{f}(p)|^2\,,
\end{equation}
we have \vspace*{-2ex}
\begin{gather}
 \bra{p} T(w^2)\ket{q}
 =\mc{F}_p\mc{T}(w^2)\mc{F}_q^{\dagger}\,,\label{Tmom}\\[1ex]
 \bra{p}G(w^2)-G_0(w^2)\ket{q}
 =\frac{\mc{F}_p}{w^2-p^2}\mc{T}(w^2)
 \frac{\mc{F}_q^{\dagger}}{w^2-q^2}\,,\label{Gmom}
\end{gather}
with elements in the matrix \eqref{Tw2} given by
\begin{gather}
 \si-\left(g,G_0(w^2)g\right)
 =\si-\int\frac{p^{1-\si}M_p}{w^2-p^2}dp\,,\label{gGg}\\[1ex]
 -\left(U_{a}g,G_0(w^2)g\right)
 =-\int\frac{e^{iap}p^{1-\si}M_p}{w^2-p^2}dp
 =i\pi w^{-\si}e^{iaw}M_w\,.\label{UGg}
\end{gather}
The integral in the last formula is calculated in the complex
plane by residues, with the use of analyticity and asymptotic
properties of $M_w$ discussed at the beginning of Appendix
\ref{P}.

\section{Reproduction of the sharp boundary
conditions}\label{odtw}

We consider now a~family of rescaled quasi-potentials $V_\la$,
$\la\in(0,1\>$, built as in (\ref{qp}), but with the use of
rescaled functions $g_\la$ instead of $g$. We write the scaling
in several equivalent forms:
\begin{equation*}
 g_\la(z)=\la^{-\frac{3}{2}}g\left(\frac{z}{\la}\right)\,,\quad
 f_\la(z)=\la^{-1-\frac{\si}{2}}f\left(\frac{z}{\la}\right)\,,\quad
 \wh{f}_\la(p)=\la^{-\frac{\si}{2}}\wh{f}(\la p)\,,
\end{equation*}
and note also that $M_{p,\la}=\la^{-\si}M_p$. The rescaled
potentials give rise to the corresponding operators
$h_{za,\la}$. All quantities referring to these operators
acquire the subscript $\la$.

Let $h_{za}^B$ be the selfadjoint, positive square root of the
operator
\begin{equation}\label{sbh}
 [h_{za}^B]^2=\left[-\frac{d^2}{dz^2}\right]_\text{boundary conditions}
\end{equation}
with  standard domains in $L^2(\mbb{R})$, Dirichlet (for
$\si=+1$)/Neumann (for $\si=-1$) conditions in $z=\pm b$, and
denote by $G^B(w^2)$ the resolvent of $[h_{za}^B]^2$. Our
objective in this section is to show the limiting property
\begin{equation} \label{odtwarzanie}
 \liml\|G_\la(w^2)-G^B(w^2)\|_\mr{HS}=0\,,
\end{equation}
where the Hilbert-Schmidt norm is
$\|A\|^2_\mr{HS}=\Tr(A^*A)\geq\|A\|^2$. From this relation it
follows then the norm convergence
\begin{equation*}
 \liml\|F(h_{za,\la})-F(h_{za}^B)\|=0
\end{equation*}
for each continuous and vanishing in infinity complex function
$F$ on $\mbb{R}$, and the strong convergence
\begin{equation*}
 \liml\|[F(h_{za,\la})-F(h_{za}^B)]\psi\|=0\,,
\end{equation*}
for each bounded continuous function $F$ and vector $\psi\in
L^2$.

It is clear from the form of Eq. \eqref{Gmom} that
$G(w^2)-G_0(w^2)$ and its scaled version $G_\la(w^2)-G_0(w^2)$
are finite rank, hence Hilbert-Schmidt, operators. Thus it is
sufficient to calculate the
\mbox{strong--$L^2(\mbb{R}^2,dp\,dq)$} limit for $\la\to0^+$ of
the integral kernel $\bra{p}G_\la(w^2)-G_0(w^2)\ket{q}$. First
we consider the numerical matrix $\mc{T}_\la$, and for later
use we also look at higher orders in $\la$. We observe that
\begin{equation}\label{granicaI0}
 \frac{1}{\la}\int\frac{M_{\la p}-M_0}{w^2-p^2}\,dp=I_0+O(\la)\,,
\end{equation}
with
\begin{equation}\label{II}
 I_0=\int\frac{M_0-M_p}{p^2}\,dp\,.
\end{equation}
This is shown by writing the difference of \eqref{granicaI0}
and \eqref{II} as
\begin{equation*}
 \la w\,\int\frac{M_u-M_0}{u^2}\,\frac{\la w}{(\la w)^2-u^2}\,du
\end{equation*}
and Fourier-transforming the integral as a scalar product of
two $L^2$-functions. This integral is shown in this way to be
bounded by a constant. Using \eqref{gGg}, \eqref{UGg},
\eqref{granicaI0} and the assumption \eqref{normN} (in Neumann
case) we find
\begin{align}
 \si-\left(g_\la,G_0(w^2)g_\la\right)
 &=i\pi(\la w)^{-\si}M_0+
 (\la w)^{1-\si}\big(\tfrac{1+\si}{2}-I_0\big)
 +O(\la^{2-\si})\,,\label{Gg}\\[1ex]
 -\left(U_{a} g_\la,G_0(w^2)g_\la\right)
 &=i\pi(\la w)^{-\si}e^{iaw}M_0+O(\la^{2-\si})\,.\label{aGg}
\end{align}
From these we get
\begin{equation*}
 \begin{split}
 \mc{T}_\la(w^2)&=\frac{-i(\la w)^\si}{\pi M_0(1-e^{2iaw})}
 \begin{pmatrix}1&-e^{iaw}\\-e^{iaw}&1\end{pmatrix}\\[1.5ex]
 &+(\la w)^{1+\si}\big(\tfrac{1+\si}{2}-I_0\big)
 \times\{\text{matrix independent of $\la$}\}+O(\la^{2+\si})\,.
 \end{split}
\end{equation*}
Next, we observe that (this is shown in Appendix \ref{P})
\begin{equation}\label{D3/2}
 \left\|\frac{\la^{\frac{\si}{2}}\wh{f}_\la(p)p^{\frac{1-\si}{2}}}{w^2-p^2}
 -\frac{\wh{f}(0)p^{\frac{1-\si}{2}}}{w^2-p^2}\right\|
 \le\begin{cases}\con(w)\, \la^{\frac{3}{2}}\,,&\text{(D)}\\
 \con(w)\, \la^{\frac{1}{2}}\,,&\text{(N)}
 \end{cases}
\end{equation}
and in addition, for Neumann case
\begin{equation}\label{N3/2}
 \left\|\frac{\la^{-\frac{1}{2}}\wh{f}_\la(p)}{w^2-p^2}
 -\frac{\wh{f}(0)}{w^2-p^2}\right\|
 \le\con(w)\, \la^{\frac{3}{2}}\q\text{(N)}
\end{equation}
(norms of functions of $p$ as elements of $L^2$). Now we easily
obtain in the $L^2(\mbb{R}^2,dp\,dq)$--sense
\begin{multline*}
 \mathrm{s}-\!\liml\bra{p}G_\la(w^2)-G_0(w^2)\ket{q}
 =-\frac{iw^{\si}}{\pi(1-e^{2iaw})}
 \frac{(pq)^{\frac{1-\si}{2}}}{(w^2-p^2)(w^2-q^2)}\\
 \times\Big[e^{-ibq} e^{ibp}-e^{iaw}e^{ibq}e^{ibp}
 +e^{ibq}e^{-ibp}-e^{iaw}e^{-ibq}e^{-ibp}\Big]\,.
\end{multline*}
We transform this to position representation and get
\begin{align*}
 &\mathrm{s}-\!\liml \bra{z}G_\la(w^2)-G_0(w^2)\ket{z'}\\
 &=-\frac{\si}{2iw\left(1-e^{2iaw}\right)}
 \bigg\{\left[\theta(b+z)e^{i(b+z)w}+\si\theta(-b-z)e^{-i(b+z)w}\right] \\
 &\hspace{1cm}\times\left[\si\theta(-b-z')e^{-i(b+z')w}+\theta(b+z')e^{i(b+z')w}\right. \\
 &\hspace{1.5cm}-\si\left.\theta(b-z')e^{iaw}e^{i(b-z')w}-\theta(-b+z')e^{iaw}e^{-i(b-z')w}\right] \\
 &\hspace{0.6cm}+\left[\theta(-b+z)e^{-i(b-z)w}+\si\theta(b-z)e^{i(b-z)w}\right] \\
 &\hspace{1cm}\times\left[\si\theta(b-z')e^{i(b-z')w}+\theta(-b+z')e^{-i(b-z')w}\right. \\
 &\hspace{1.5cm}-\si\left.\theta(-b-z')e^{iaw}e^{-i(b+z')w}-\theta(b+z')e^{iaw}e^{i(b+z')w}\right]\bigg\}\,.
\end{align*}
We shall use also the explicit form of the unperturbed Green function in this
representation
\begin{equation}\label{G0}
 \bra{z}G_0(w^2)\ket{z'}=-\frac{i}{2w}\left[\theta(z-z')e^{i(z-z')w}
 + \theta(z'-z) e^{i(z'-z)w} \right] \,.
\end{equation}
In this way we find
\begin{equation*}
 \mathrm{s}-\!\liml\bra{z}G_\la(w^2)\ket{z'}=\bra{z}G^B(w^2)\ket{z'}\,,
\end{equation*}
where
\begin{multline}\label{GB}
 \bra{z}G^B(w^2)\ket{z'}
 =\left[\bra{z}G_0(w^2)\ket{z'}+\si\frac{i}{2w}e^{-i(z+z'+a)w}\right]
 \chi_{(-\infty,-b)}(z)\chi_{(-\infty,-b)}(z')\\
 + \left[\bra{z}G_0(w^2)\ket{z'}+\si\frac{i}{2w} e^{i(z+z'-a)w}
 \right] \chi_{(b,+\infty)}(z)\chi_{(b,+\infty)}(z')\\
 + \bigg[\bra{z}G_0(w^2)\ket{z'}+\frac{i}{w}
 \bigg(\frac{\cos(zw)\cos(z'w)}{1+\si e^{-iaw}}
 +\frac{\sin(zw)\sin(z'w)}{1-\si e^{-iaw}}\bigg)\bigg]
 \chi_{(-b,b)}(z)\chi_{(-b,b)}(z')\,,
\end{multline}
and $\chi_\Omega$ is the characteristic function of the set
 $\Omega$.

In the three regions $\bra{z}G^B(w^2)\ket{z'}$ differs from
$\bra{z}G_0(w^2)\ket{z'}$ only by solutions of homogeneous
equation and satisfies the boundary conditions
\begin{align*}
 \bra{\pm b}G^B(w^2)\ket{z'}&=\bra{z}G^B(w^2)\ket{\pm b}=0\,,\quad &&\text{(D)}\\
 \frac{d}{dz}\bra{z}G^B(w^2)\ket{z'}|_{z=\pm b}
 &=\frac{d}{dz'}\bra{z}G^B(w^2)\ket{z'}|_{z'=\pm b}=0\,,\quad &&\text{(N)}
\end{align*}
so it is indeed the Green function of the Dirichlet/Neumann
operator and therefore $\eqref{odtwarzanie}$ is finally proven.

We now want to acquire some information on the rate at which
the limit (\ref{odtwarzanie}) is achieved. Using \eqref{D3/2}
one finds for any $\vp,\eta\in L^2$
\begin{equation}\label{weakDN}
 \left(\vp,G_\la(w^2)\eta\right)
 =\left(\vp,G^B(w^2)\eta\right)+
 \begin{cases}(1-I_0)\,O(\la)+O(\la^{3/2})\,,&\text{(D)}\\
 O(\la^{1/2})\,.&\text{(N)}
 \end{cases}
\end{equation}
The Neumann case turns out to be here, as in many other
problems, more singular. However, we also note that if we
assume that $\vp$ and $\eta$ are in the domain of $h_z$ then
the estimate \eqref{N3/2} implies
\begin{equation}\label{weakN}
 \left(\vp,G_\la(w^2)\eta\right)
 =\left(\vp,G^B(w^2)\eta\right)
 +I_0O(\la)+O(\la^{3/2})\,.\qquad\text{(N)}
\end{equation}

In the following two sections we treat unscaled models. The
scaling is again considered in Section \ref{scaling}.

\section{Spectral analysis}
\label{spectral}

We add now some further assumptions on the choice of functions
$f$. We denote for $k\in\mbb{R}$
\begin{equation} \label{Ik}
 I_k = \int \frac{M_k-M_p}{p^2 - k^2} dp \,,
\end{equation}
and demand that
\begin{gather}
 0<I_k\quad\text{for}\ k\neq0\quad\text{(D,N)}\,,\label{Irange}\\
 I_k<1\quad\text{(D)}\,.\label{IrangeD}
\end{gather}
Note that by continuity $I_0\geq0$ (this is the quantity
introduced in (\ref{II})). We also denote
\begin{equation} \label{Jk}
 \pi N_{k}=|k|^\si \Big\{\si + \int \frac{q^{1-\si} M(q) -
 k^{1-\si} M(k)}{q^2-k^2} dq\Big\} 	
 =\tfrac{1}{2}(1+\si)|k|^\si-|k|I_k \,.
\end{equation}

The operators $h_{za}^2$ are nonnegative, and outside a compact
set in $\mbb{R}$ they act as $-\partial_z^2$. Therefore their
continuous spectrum covers the whole positive axis and thus the
spectrum is $\<0,+\infty)$. This does not resolve the question
of point spectrum, and we treat it first.

The eigenvector equation
\begin{equation}\label{eigen}
 h_{za}^2 \psi_{k} = k^2 \psi_{k} \,,\quad
 \psi_{k}\in L^2(\mbb{R})\,,\quad k\geq0
\end{equation}
is solved in momentum space. It is easily seen that the
distributional solution which is square-integrable at infinity
must have the form
\begin{equation} \label{st_zw}
 \wh{\psi}_{k}(p)=\frac{c_{b} e^{-ibp}
 + c_{-b} e^{ibp}}{p^2 - k^2} p^{\frac{1-\si}{2}} \wh{f} (p)\,,
\end{equation}
with constants $c_{\pm b}$ to be determined. Putting this form
back into Eq. \eqref{eigen} one finds for $k>0$ that the
constants $c_{\pm b}$ have to satisfy the linear system
\begin{equation}\label{rownanie_na_c}
 N_k\,c_{+b}-M_k \sin(ak)\,c_{-b}=0\,,\quad
 M_k \sin(ak)\,c_{+b}-N_k\,c_{-b}=0\,,
\end{equation}
where the integration leading to coefficients $M_k \sin(ak)$ is
carried out with the use of analyticity and asymptotic behavior
of $M_k$ discussed in Appendix \ref{P}. Now, for $k>0$ the
conditions \eqref{Irange} and \eqref{IrangeD} imply $N_k\neq0$,
thus nontrivial solutions to the system $\eqref{rownanie_na_c}$
exist only if \mbox{$M_k \sin(ak)=\pm N_k$}, and in that cases
$\eqref{st_zw}$ takes, respectively, the form
\begin{equation*}
 \wh{\psi}_{k}(p)=c_{-b}\frac{e^{ibp}\mp e^{-ibp}}{p^2-k^2}
 p^{\frac{1-\si}{2}} \wh{f}(p)\,.
\end{equation*}
The condition $\wh{\psi}_{k}\in L^2(\mbb{R})$ requires that
 $\wh{f}(\pm k) = 0$ or $e^{ibk}\pm e^{-ibk} = 0$.
Each of these cases implies $M_k\sin(ak)=0$, which shows that
there are no eigenvectors for $k>0$.

For $k=0$ in Dirichlet case the solution \eqref{st_zw} cannot
be in $L^2$ as $\wh{f}(0)\neq0$. For Neumann case one finds
that \eqref{st_zw} is a distributional solution for any
constants $c_{\pm b}$, so we are free to choose them so as to
satisfy the square-integrability of \eqref{st_zw}. This happens
only for $c_{-b}=-c_b$ and then
\begin{equation} \label{psi0}
 \wh{\psi}_{0}(p)=\mc{N}\,\frac{\sin(bp)}{p}\,\wh{f}(p)\,, \qquad \text{(N)}
\end{equation}
where $\mc{N}$ is a proportionality factor. The normalization
condition $||\wh{\psi}_0||_{L^2}=1$ gives
\begin{equation} \label{psi0N}
 |\mc{N}|^2 = \frac{2}{a\pi M_0-I_0} \,.
\end{equation}
Summarizing, there are no bound states for Dirichlet case,
however for  Neumann case there is one bound state, which
corresponds to the zero eigenvalue,  described by \eqref{psi0}
and \eqref{psi0N}.

We now consider the continuous spectrum and for this purpose
use the stationary scattering formalism. The improper
eigenfunctions of scattering states in momentum representation
are given in standard notation by
\begin{equation} \label{psi_scatt}
 \wh{\psi}_{k}(p)= \braket{p}{k+} = \delta(p-k) + \frac{\bra{p} T(k^2+i0)
 \ket{k}}{k^2-p^2+i0} \,,
\end{equation}
where $T(w^2)$ is the operator discussed in Section
\ref{models}. The variable $k$ takes all values $k\neq0$ and
each spectrum point $k^2$ has two-fold degeneracy corresponding
to $\pm k$. Taking into account the results of Section
\ref{models} we can write
\begin{equation}\label{TMJ}
 \mc{T}(k^2+i0) =
 \frac{|k|^\si}{i\pi}\begin{pmatrix}
 M_k-iN_k&M_k\,e^{ia|k|}\\[1.5ex]
 M_k\, e^{ia|k|}&M_k-iN_{k}
 \end{pmatrix}^{-1}\,.
\end{equation}
For later use we write this in two alternative forms. We
introduce matrix notation
\begin{gather}
 \mc{M}(k) = \mc{F}_k^{\dagger}\mc{F}^{}_k\,,\quad
 \mc{N}(k)=\frac{2|k|}{\pi}
 \left[\si\mathbbm{1}+\mc{P}\int\frac{\mc{M}(p)}{p^2 - k^2} dp
 \right]\,,\label{MN}\\[1ex]
 \mc{L}(k)= \mc{M}(k)+\mc{M}(-k)=2 k^{1-\si} M_k
 \begin{pmatrix}
 1 & \cos(ak)\\
 \cos(ak) & 1 \\
 \end{pmatrix} \,.\label{L}
\end{gather}
and then
\begin{gather}
\mc{T}(k^2+i0)=\frac{2|k|}{\pi}\big(\mc{N}(k)+i\mc{L}(k)\big)^{-1}\,,\label{TL}\\[1ex]
 \left| \bra{p}T(k^2+i0)\ket{k}\right|^2
 =\Tr\left[\mc{M}(p)\mc{T}(k^2+i0)\mc{M}(k)\mc{T}(k^2+i0)^{\dagger}\right]\,.\label{TM}
\end{gather}
Finally, we calculate the inverse in \eqref{TMJ} and write the
result in the form
\begin{equation}\label{TMs}
 \mc{T}(k^2+i0) =\frac{1}{i\pi}\,
 \frac{|k|^\si s_k}{1-(q_ke^{ia|k|})^2}
 \begin{pmatrix}1&-q_ke^{ia|k|}\\[1.5ex]
 -q_ke^{ia|k|}&1\end{pmatrix}\,,
\end{equation}
where
\begin{equation*}
 s_k=\frac{1}{M_k-iN_k}\,,\quad q_k=\frac{M_k}{M_k-iN_k}=M_ks_k\,.
\end{equation*}
Some properties of $s_k$ function are shown in Appendix
\ref{P}.

\section{Hilbert-Schmidt properties}
\label{HS}

In this section we show that our model satisfies the
admissibility conditions \eqref{HS1} and \eqref{HS2}. If we
write $\mr{TR}_\tau$ for the l.h.s. of these two conditions,
with $\tau=0$ for \eqref{HS1} and $\tau=1$ for \eqref{HS2},
then
\begin{equation}\label{TR}
 \mr{TR}_\tau=
 \int_{\mbb{R}^2}|p|^\tau|\bra{p}h_{za}-h_z\ket{k+}|^2dk\,dp
 +\tfrac{1-\si}{2}\int_{\mbb{R}}|p|^{2+\tau}\big|\wh{\psi}_{0}(p)\big|^2dp\,,
\end{equation}
where the first term results from the continuous spectrum space
of $h_{za}$ and the second is the bound state contribution in
Neumann case. The second term is evidently finite (by
\eqref{psi0}), and we restrict attention to the first one,
which we denote $\mr{TR}_\tau^\mr{cont}$. In momentum
representation we have
\begin{equation*}
 \bra{p}h_{za}-h_z\ket{k+}=(|k|-|p|)\wh{\psi}_{k}(p)
 =\frac{\bra{p} T(k^2+i0)\ket{k}}{|p|+|k|}\,,
\end{equation*}
thus by change of variables for negative arguments we get
\begin{equation}\label{TRcont}
 \mr{TR}_\tau^\mr{cont}=\int_{\mbb{R}_+^2}
 \frac{p^\tau}{(p+k)^2}
 \sum_{\pm\pm}\big|\bra{\pm p}T(k^2+i0)\ket{\pm k}\big|^2 dk\,dp\,,
\end{equation}
where the signs in `bra' and `ket' are uncorrelated and the sum
is over all four possibilities. From now on we assume that
$k,p\ge0$. Using Eqs.\ \eqref{L}, \eqref{TM} we can write
\begin{equation*}
 \sum_{\pm\pm}\big|\bra{\pm p}T(k^2+i0)\ket{\pm k}\big|^2
 =\Tr\left[ \mc{L}(p)\mc{T}(k^2+i0)\mc{L}(k)\mc{T}(k^2+i0)^{\dagger}\right]\,,
\end{equation*}
and substituting here (from \eqref{TL})
\begin{equation*}
 \mc{L}(k)=\frac{ik}{\pi}
 \left(\mc{T}(k^2+i0)^{\dagger-1}-\mc{T}(k^2+i0)^{-1}\right)
\end{equation*}
we give Eq.\ \eqref{TRcont} the form
\begin{equation} \label{trace2}
 \mr{TR}_\tau^\mr{cont}=\frac{2}{\pi}
 \int_{\mbb{R}_+^2}
 \frac{kp^\tau}{(p+k)^2}\,\Realis\Big[
 i\Tr\left[\mc{L}(p)\mc{T}(k^2+i0)\right]\Big]dk \, dp\,.
\end{equation}
With the use of \eqref{L} and \eqref{TMs} we have
\begin{equation}\label{kRTr}
 \begin{split}
 &k\Realis\Big[i\Tr\left[\mc{L}(p)\mc{T}(k^2+i0)\right]\Big]=
 \frac{4}{\pi}\,p^{1-\si}k^{1+\si}M_p\Realis\bigg[s_k\,
 \frac{1-\cos(ap)q_ke^{iak}}{1-(q_ke^{iak})^2}\bigg]
\end{split}
\end{equation}
Writing out the real part gives us the appropriate behavior of
the nominator for $k=0$ and the whole expression becomes
proportional to $M_k$. Using the estimates \eqref{est_denom},
\eqref{s_k} and \eqref{estIk} one finds that
$\mr{TR}_\tau^\mr{cont}$ are finite and the admissibility
conditions \eqref{HS1} and \eqref{HS2} are satisfied.

\section{Scaling}
\label{scaling}

We return to the scaling transformation to view it from a
different point. If we make the dependence of the potential $V$
on $a$ explicit by writing it as $V_a$ and the rescaled
potential as $V_{a,\la}$ then we have
\begin{equation*}
 V_{a,\la}(z,z')=\la^{-3}V_{a/\la}(z/\la,z'/\la)\,.
\end{equation*}
It is then an easy exercise to show that this implies a simple
scaling law of the eigenfunctions \eqref{psi0} and
\eqref{psi_scatt}:
\begin{equation*}
 \wh{\psi}_{k,b}^\la(p)=\la\wh{\psi}_{\la k,b/\la}(\la p)\,,\quad
 \wh{\psi}_{0,b}^\la(p)=\la^{1/2}\wh{\psi}_{0,b/\la}(\la p)
\end{equation*}
(different powers of $\la$ reflect different normalizations: to
the Dirac delta and to unity respectively).  Denoting the
scaled versions of \eqref{TR}, with explicit dependence on $a$,
by $\mr{TR}_{\tau,a}^\la$ we find
\begin{equation*}
 \mr{TR}_{\tau,a}^\la=\la^{-2-\tau}\,\mr{TR}_{\tau,a/\la}\,.
\end{equation*}
Thus the admissibility conditions are satisfied also for the
rescaled potential (but not in the limit). In the same way one
obtains the scaling law for the Casimir energy:
\begin{equation}\label{en-scale}
 \ve_a^\la=\la^{-3}\,\ve_{a/\la}\,.
\end{equation}
Therefore to identify the scaling behavior of the energy in the
limit it is sufficient to expand the unscaled energy in inverse
powers of $a$ up to the third order. This will be done in the
next section.

\section{The energy}\label{enexp}

In this section we prove the following expansion of Casimir
energy for large $a$:
\begin{align}
 \ve_a&=\ve_{\infty}-\frac{\pi^2}{1440 a^3}+o(a^{-3})\,,\quad
 &&\text{(D)}\label{CED}\\
 \ve_a&=\ve_{\infty}+\frac{1}{48\pi M_0 a^2}
 +\frac{I_0}{8\pi^2 M_0^2 a^3}\left(\frac{\zeta(3)}{\pi^2}+\frac{1}{3}\right)
 -\frac{\pi^2}{1440 a^3}+o(a^{-3})\,.&&\text{(N)}\label{CEN}
\end{align}
We postpone the discussion of this result to the concluding
section and here only note that there are functions in our
class for which $I_0=0$, and then the $a^{-3}$-term has the
known universal form.

The two conditions \eqref{HS1} and \eqref{HS2}, considered in
Section \ref{HS}, imply already finiteness of the Casimir
energy per unit area \eqref{main_energy}, as mentioned in the
Introduction. This can be easily seen: we observe that
conditions \eqref{HS1} and \eqref{HS2} mean that
$\Delta=h_{za}-h_z$ and $h_z^{1/2}\Delta$ are Hilbert-Schmidt
operators. Also, from
\begin{equation*}
 \Delta h_{za}\Delta=\Delta h_z\Delta+\Delta^3
\end{equation*}
we infer that $h_{za}^{1/2}\Delta$ is HS as well, which is
sufficient for the claim.

The expression \eqref{main_energy} closely parallels that of
the condition \eqref{HS2} and can be written in analogy to
\eqref{TR}. We split the trace in \eqref{main_energy} into
two terms and calculate these traces in $h_{za}$ or $h_z$
(improper) basis in the first and second line below, respectively.
Then we insert the spectral decomposition of the operators $h_z$
or $h_{za}$, respectively. In this way we get
\begin{gather*}
 \Tr\,\big[2\Delta h_z \Delta\big]=2\int_{\mbb{R}^2}
 |p|\, \big|\bra{p}h_{za}-h_z\ket{k+}\big|^2 dk\,dp
 +(1-\si)\int_{\mbb{R}}|p|^3\big|\wh{\psi}_{0}(p)\big|^2dp\,,\\
 \Tr\,\big[\Delta h_{za} \Delta\big]=\int_{\mbb{R}^2}
 |k|\, \big|\bra{p}h_{za}-h_z\ket{k+}\big|^2 dk\,dp\,.
\end{gather*}
Therefore the expression \eqref{main_energy} may be written in the form
\begin{equation}\label{energy_main}
 \ve_a=\frac{1}{24\pi}
 \int_{\mbb{R}^2}(2|p|+|k|)\big|\bra{p}h_{za}-h_z\ket{k+}\big|^2dk\,dp
 +\frac{1-\si}{24\pi}\int_{\mbb{R}}|p|^{3}\big|\wh{\psi}_{0}(p)\big|^2dp\,.
\end{equation}
The bound state contribution $\ve_a^\mr{bound}$ -- the second
term -- will be calculated in the Neumann case subsection.
Following the same steps as in Section \ref{HS} and using
\eqref{kRTr} we give the continuous spectrum contribution the
form
\begin{equation} \label{energia}
 \ve_{a}^\mr{cont}=\frac{1}{3\pi^3}\int_{\mbb{R}_+^2}\chi(k,p)M_p
 \Realis\bigg[s_k\,\frac{1-\cos(ap)q_ke^{iak}}{1-(q_ke^{iak})^2}\bigg]
 dk\,dp\,,
\end{equation}
where
\begin{equation*}
 \chi (k,p)=\frac{k^{1+\si}p^{1-\si}(2p+k)}{(p+k)^2}\,.
\end{equation*}
We write this as the limit for $\ep\to0^+$ of the integral
restricted to $k\in\<\ep,+\infty)$ and expand the denominator
into geometric power series (note that $|q_k|<1$ for $k>0$)
\begin{equation*}
 \frac{1}{1-\left(q_k e^{iak}\right)^2}
 =\sum_{n=0}^\infty(q_ke^{iak})^{2n}\,,
\end{equation*}
getting
\begin{multline*}
 \ve_{a}^\mr{cont}=\frac{1}{3\pi^3} \lime \int_0^\infty
 M_p \int_\epsilon^\infty \chi (k,p)
 \Realis\bigg[s_k \sum_{n\in 2\mbb{N}_0}q_k^n e^{inak} \\
 -s_k \cos (ap) \sum_{n\in2\mbb{N}-1}q_k^n e^{inak} \bigg]dk \, dp\,.
\end{multline*}
We write the $n=0$ term separately as
\begin{equation} \label{energia0}
 \ve_{\infty}
 =\frac{1}{3\pi^3}\int_{\mbb{R}_+^2}
 \chi(k,p)M_p M_k |s_{k}|^2 dk \,dp\,.
\end{equation}
For other terms we observe that
\begin{equation}\label{sMest}
 \sum_{n=1}^N|q_ke^{iak}|^{2n}\leq
 \frac{|s_k|^2M_k^2}{1-|q_k|^2}\leq\mr{const}\,M_k^2\,\frac{1+k^{m-2\si}}{k^m}
 \quad
 \begin{cases}
 m=2&\text{(D)}\\
 m\geq2&\text{(N)}
 \end{cases}
\end{equation}
(see \eqref{estden2} and \eqref{s_k}), so we can pull the
infinite sum sign outside the integral by the Lebesgue
dominated convergence theorem to obtain
\begin{equation} \label{energia2}
 \begin{split}
 \ve_a^\mr{cont}=\ve_{\infty}
 +\frac{1}{6\pi^3}\lime\sum_{n\in 2\mbb{N}}\int_0^{\infty}
 M_p\left[\int_{\epsilon}^{\infty}\chi(k,p)s_k\,(q_k e^{iak})^n dk
 + \ \tr{c.c.}\right] dp&\\
 -\frac{1}{6\pi^3}\lime\sum_{n\in 2\mbb{N}-1}\int_0^{\infty}
 M_p\cos(ap)\left[\int_{\ep}^{\infty}\chi(k,p)s_k(q_k e^{iak})^n dk
 + \ \tr{c.c.} \right] dp&\,.
 \end{split}
\end{equation}
We now split the analysis into separate cases.

\subsection{Dirichlet case ($\si=1$)}\label{enexpD}

In this case when \eqref{sMest} is multiplied by $\chi(k,p)$
the $k^{-2}$ singularity in $k=0$ is canceled by $k^2$ from the
function $\chi$. Therefore here the intermediate step with
nonzero $\ep$ is not needed and \eqref{energia2} should be read
with $\ep=0$. Moreover, there is no bound state here, so this
formula represents the total Casimir energy $\ve_a$. If we
write $(na)^3e^{inak}=(-i\partial_k)^3e^{inak}$ and integrate
three times by parts we find
\begin{equation*}
 (na)^3\int_0^{\infty} \chi(k,p) s_k\,(q_k e^{iak})^n dk
 = - i\,\frac{4M_p}{M_0 p}-
 i\int_0^{\infty}\partial_k^3 \Bigl(\chi(k,p)s_k\,q_k^n \Bigr)e^{inak} dk\,,
\end{equation*}
and then obtain (the first term on the r.h.s. above is
imaginary and falls out)
\begin{multline} \label{energyD}
 \ve_{a}-\ve_{\infty}=\frac{i}{6\pi^3a^3}
 \Bigg\{-\sum_{n\in2\mbb{N}}\frac{1}{n^3}
 \int_{\mbb{R}_+^2}M_p\partial_k^3
 \Bigl(\chi(k,p)s_k\,q_k^n \Bigr)e^{inak}dk\,dp\\
 +\sum_{n\in2\mbb{N}-1}\frac{1}{n^3}
 \int_{\mbb{R}_+^2}M_p\cos(ap)\partial_k^3
 \Big(\chi(k,p)s_k\,q_k^n \Bigr)e^{inak}dk\,dp\Bigg\} \
 + \ \tr{c.c.} \,.
\end{multline}
Let now $\Omega$ be the intersection of $\mbb{R}_+^2$ with an
arbitrary neighborhood of zero. We now use the results of
Appendix \ref{chi'''} to infer that the following successive
three operations on this formula lead only to the neglect of
terms of order $o(a^{-3})$:
\begin{itemize}
 \item[(i)] replacement of \label{strona(i)}
     $\partial_k^3\big(\chi(k,p)s_kq_k^n \big)$ by
$s_kq_k^n \partial_k^3\chi(k,p)$;
 \item[(ii)] restriction of the integration region to
     $\Omega$;
 \item[(iii)] replacement (in the restricted region) of
     $M_p s_kq_k^n $ by $M_0s_0q_0^n=1$.
\end{itemize}
In this way we arrive at
\begin{multline}\label{enom}
 \ve_{a}= \ve_{\infty} + \frac{2}{\pi^3 a^3} \sum_{n \in 2 \mbb{N}}
 \frac{1}{n^3} \int_{\Omega}
 \frac{p^2(k-3p)}{(k+p)^5} \sin(nak) dk \, dp \\
 - \frac{2}{\pi^3 a^3} \sum_{n \in 2 \mbb{N}-1} \frac{1}{n^3}
 \int_{\Omega}
 \frac{p^2(k-3p)}{(k+p)^5} \cos(ap)\sin(nak) dk \, dp + \: o
 \Bigl( \frac{1}{a^3} \Bigr)\,.
\end{multline}
The integrals are bounded uniformly with respect to $a$. It is
now sufficient to show that they have well defined limits for
$a\to\infty$; then those limits determine the $a^{-3}$-term.

Consider the integrals (with $k,p>0$)
\begin{equation*}
 C(n,\ell,a)=\int_{k+p\leq1}\left( \frac{p^2}{(k+p)^4} -
 \frac{4p^3}{(k+p)^5} \right)\sin(nak+\ell ap)
 dp\,dk\,,\qquad \ell=0,\pm1\,.
\end{equation*}
It is easy to show that with the choice
$\Omega=\{k,p>0,k+p\leq1\}$ the integral in the first line of
\eqref{enom} is this integral with $\ell=0$, while the integral
in the second line of \eqref{enom} is one half of the sum of
integrals with $\ell=\pm1$. By the change of variables
$r=a(k+p)$, $t=p/(k+p)$ we bring this to the form
\begin{equation*}
 C(n,\ell,a)= \int_0^1 \left(t^2 - 4t^3\right)\int_0^a
 \frac{1}{r}\sin\left[r\bigl(n+t(\ell-n)\bigr)\right]dr\,dt\,,
\end{equation*}
and find
\begin{equation*}
 \lim_{a\to\infty}C(n,\ell,a)=
 \begin{cases}
 -\frac{\pi}{3} & \tr{for $\ell=0,1$}, \\
 \frac{\pi}{3} \left( 1 + \frac{n^3}{(n+1)^3} - \frac{3
 n^4}{(n+1)^4} \right) & \tr{for $\ell=-1$}.
 \end{cases}
\end{equation*}
Using these results we get finally
\begin{multline*}
 \ve_{a} = \ve_{\infty} - \frac{1}{3 \pi^2 a^3}
 \left[\sum_{n\in2\mbb{N}}\frac{2}{n^3}
 + \sum_{n \in 2 \mbb{N}-1}
 \left(\frac{1}{(n+1)^3} - \frac{3n}{(n+1)^4} \right)\right]
 + \: o \Bigl( \frac{1}{a^3} \Bigr)\\
 = \ve_{\infty} - \frac{\zeta(4)}{16 \pi^2 a^3} + \:
 o\Bigl(\frac{1}{a^3}\Bigr)=\ve_{\infty}-\frac{\pi^2}{1440 a^3}
 + \: o \Bigl( \frac{1}{a^3} \Bigr)\,,
\end{multline*}
where $\ve_{\infty}$ is defined in $\eqref{energia0}$. 	

\subsection{Neumann case ($\si=-1$)}\label{enexpN}

We now use formula $\eqref{energia2}$ with the replacement in
notation $q_k e^{iak} = \tilde{q}_k e^{i\tilde{a}k}$, where
\begin{equation*}
 \alpha = \frac{I_0}{\pi M_0}, \q
 \tilde{a}=a-\alpha, \q \tilde{q}_k=q_k e^{i\alpha k}\,,
\end{equation*}
which has the technical advantage that $\tilde{q}'_0=0$. After
this modification the general scheme is very similar to the
Dirichlet case. Integration by parts gives expansion in
$1/\tilde{a}$ but at the end we shall translate it to the $1/a$
expansion. Integrating by parts we obtain boundary terms, for
which in the present case the addition of c.c. terms must be
taken into account before the limit $\ep\to0^+$ is performed.
For example, the first integration by parts in $k$ in the first
line in $\eqref{energia2}$ gives a term proportional to (before
$p$-integration)
\begin{multline*}
 \lime \sum_{n \in 2 \mbb{N}} \frac{1}{n} \chi (\ep, p)
 \Imaginalis\left[s_k \tilde{q}_k^n (\ep)
 e^{in\tilde{a}\ep}\right]\\[-1.5ex]
 = - \lime \chi (\ep,p)
 \Imaginalis\big[ s_{\ep}
 \ln\left(1-\tilde{q}_\ep^2 e^{2i\tilde{a}\ep} \right)\big]
 = \frac{\pi p}{M_0} \,,
\end{multline*}
where in the last equality we used the fact that in the
neighborhood of zero $\left|\ln|1 - \tilde{q}_\ep^2
e^{2i\tilde{a}\ep}|\right|\le\ln\frac{\con(a)}{\ep}$  (see
\eqref{est_denom}). After integrating by parts three times in
similar way we get the expression
\begin{multline*}
 \lime \left[ \sum_{n \in 2 \mbb{N}} \int_{\epsilon}^{\infty}  \chi(k,p)
 s_k\tilde{q}_k^n e^{in\tilde{a}k} dk  + \ \tr{c.c.} \right]
 = - \frac{\pi p}{\tilde{a} M_0 } + \frac{\pi^2}{4\tilde{a}^2 M_0}
 + \frac{3 I_0 \zeta(3)}{2\tilde{a}^3 \pi M_0^2}  \\
 - \lime \left[ \sum_{n \in 2 \mbb{N}} \frac{1}{(in\tilde{a})^3}
 \int_{\epsilon}^{\infty}  \partial_k^3 \Big( \chi(k,p) s_k\tilde{q}_k^n
 \Big) e^{in\tilde{a}k} dk
 + \ \tr{c.c.} \right] \,.
\end{multline*}
For the sum over odd natural numbers the computations are
similar. The integrals over $p$ we treat in a similar way as in
the Dirichlet case (steps (i) to (iii), for their
permissibility here see Appendix \ref{chi'''}). The integrals
of $\partial_k^3 \chi$ over the neighborhood of zero go
similarly as in the Dirichlet case. In this way we find
\begin{multline*}
 \ve_{a}^\mr{cont}=\ve_{\infty}-\frac{C_M}{6\pi^2 M_0 \tilde{a}}
 + \frac{1}{48 \pi M_0 \tilde{a}^2} \\
 - \frac{1}{\tilde{a}^3} \left[ \frac{1}{6 \pi^2}
 - \frac{I_0 \zeta(3)}{8\pi^4 M_0^2}
 + \frac{\pi^2}{1440}\right]
 + \: o \Bigl( \frac{1}{\tilde{a}^3} \Bigr)\,,
\end{multline*}
where
\begin{equation*}
 C_M = \int_0^\infty p M_p dp\,.
\end{equation*}
In the calculation we used the relations
\begin{equation} \label{pomocne_wzory}
 \int_0^\infty M_p dp = \frac{1}{2} \,, \qquad
 \int_0^\infty p M_p \cos(ap) dp
 = - \frac{M_0}{\tilde{a}^2} + \: O \Bigl( \frac{1}{\tilde{a}^3} \Bigr)\,.
\end{equation}
The first equality follows from normalization of $f$. We now
return to the bound state contribution to the energy. From the
second term of \eqref{energy_main}, \eqref{psi0} and
\eqref{psi0N}, using \eqref{pomocne_wzory} we get
\begin{equation*}
 \ve_{a}^\mr{bound}
 = \frac{C_M}{6\pi^2 M_0 \tilde{a}}
 + \frac{1}{6\pi^2 \tilde{a}^3}
 + \: o \Bigl( \frac{1}{\tilde{a}^3} \Bigr)\,.
\end{equation*}
Adding the two contributions and changing the expansion
parameter to $a$ we obtain~\eqref{CEN}.

\section{Local properties}\label{local}

In this section we consider the local algebras of fields
supported outside the regions of support of the potential $V$.
For the initial (unscaled) models this means that the
$z$-support of fields is outside the set
$\<-b-R,-b+R\>\cup\<b-R,b+R\>$, but for $\la\to0$ eventually
every support outside the planes $z=\pm b$ is admitted. Fields
thus supported are also in the algebra of fields of the models
with sharp boundary conditions at $z=\pm b$, so one can also
consider sharp boundary conditions for them.

We recall from \cite{her05} that we use the initial value
fields (smeared on a Cauchy surface of constant time)
$\Phi(V)$, where $V$ is a pair of real test functions $(v,u)$,
and $X(u)=\Phi(0,u)$, $P(v)=\Phi(v,0)$ have the interpretation
of canonical variables. For the present choice of the algebras
the test functions are assumed to be in the space of smooth
functions of compact support outside $z=\pm b$, which we denote
by $\mc{D}_b$. The algebra of fields is formulated, more
precisely, in Weyl form, which means that rather than $\Phi(V)$
elements $W(V)=\exp[i\Phi(V)]$ are used, and the fields $\Phi$
are defined on the level of specific representations. The
states on the algebra are given as normalized positive linear
functionals on the algebra of Weyl elements.

With the Hamiltonians of the perpendicular motion $h_{za}$ and
$h_{za}^B$ as defined in \eqref{hqp} and \eqref{sbh} we denote
\begin{equation}
 h_a^2=h_{za}^2+h_\perp^2\,,\quad
 [h_a^B]^2=[h_{za}^B]^2+h_\perp^2\,,
\end{equation}
where $-h_\perp^2$ is the two-dimensional Laplacian in the
directions parallel to the plates. Then the ground states of
the fields corresponding to the models proposed in the present
article and to the sharp boundary conditions are given,
respectively, by
\begin{equation}
 \omega_a\big(W(V)\big)=\exp\big[-\tfrac{1}{4}\|j_a(V)\|^2\big]\,,\quad
 \omega_a^B\big(W(V)\big)=\exp\big[-\tfrac{1}{4}\|j_a^B(V)\|^2\big]\,,
\end{equation}
where
\begin{equation}
 j_a(V)=h_a^{1/2}v-ih_a^{-1/2}u\,,\quad
 j_a^B(V)=\big[h_a^B\big]^{1/2}v-i\big[h_a^B\big]^{-1/2}u\,.
\end{equation}
(To be precise, to obtain formula \eqref{main_energy} for the
Casimir energy we start in \cite{her06} with the directions
parallel to the plates restricted to a box, whose size then
tends to infinity. This may be shown to reproduce the states
given above, but we omit this step here.) The ground states
$\omega_{a,\la}$ of the scaled models are defined analogously
with the use of $h_{a,\la}$.

We show in this section that the scaled states reproduce in the
weak limit the sharp boundary state:
\begin{equation}\label{limstat}
 \lim_{\la\to0}\omega_{a,\la}\big(W(V)\big)
 =\omega_a^B\big(W(V)\big)\,.
\end{equation}
Also, we consider the limit of the local energy density.

\subsection{Local limit of states}

There is $\|j_a(V)\|^2=(v,h_av)+(u,h_a^{-1}u)$, thus to prove
\eqref{limstat} it is sufficient to show that
$(\vp,h_{a,\la}^{\pm1}\psi)$ tend to
$(\vp,\big[h_a^B\big]^{\pm1}\psi)$ for $\la\to0$ and
$\vp,\psi\in\mc{D}_b$. For such $\psi$ one has
\begin{equation*}
 h_{a,\la}^2\psi=\big[h_a^B\big]^2\psi=h^2\psi
 \equiv-\Delta\psi
\end{equation*}
for sufficiently small $\la$. Therefore
$(\vp,h_{a,\la}\psi)=-(\vp,h_{a,\la}^{-1}\Delta\psi)$ and
similarly for $h_{a}^B$, and the problem is reduced to the
$h_{a,\la}^{-1}$-case. Moreover, $\vp$ and $\psi$ are in the
domain of $h_\perp^{-1/2}$, so we can write
 $(\vp,h_{a,\la}^{-1}\psi)
 =\big(h_\perp^{-1/2}\vp,h_\perp^{1/2}h_{a,\la}^{-1}h_\perp^{1/2}
(h_\perp^{-1/2}\psi)\big)$ and similarly for $h_a^B$. In this
way the problem is reduced to the following:
\begin{equation}\label{wlimlab}
 \mathrm{w}-\lim_{\la\to0}h_\perp^{1/2}h_{a,\la}^{-1}h_\perp^{1/2}
 =h_\perp^{1/2}\big[h_a^B\big]^{-1}h_\perp^{1/2}\,.
\end{equation}
To show this we first observe that
$\|h_\perp^{1/2}h_{a,\la}^{-1}h_\perp^{1/2}\|\leq1$, so it is
sufficient to perform the limit between vectors from the total
set of the form
$\chi(\vec{x})=\chi_\perp(\vec{x}_\perp)\chi_z(x^3)$. Using the
spectral representation of $h_\perp^2$ one has
\begin{equation}
 (\chi,h_\perp^{1/2}h_{a,\la}^{-1}h_\perp^{1/2}\rho)
 =\int\bigg(\chi_z,
 \frac{|\vec{p}_\perp|}{\sqrt{h_{za,\la}^2+\vec{p}_\perp^{\,\,2}}}\,
 \rho_z\bigg)\,
 \ol{\wh{\chi_\perp}(\vec{p}_\perp)}\wh{\rho_\perp}(\vec{p}_\perp)\,d^2p_\perp\,.
\end{equation}
The scalar product under the integral is bounded by
$\|\chi_z\|\|\rho_z\|$ and for each $\vec{p}_\perp$, by the
result of Section \ref{odtw}, tends to analogous expression
with $h_{za,\la}$ replaced by $h_{za}^B$. This is sufficient to
perform the limit, which ends the proof.

\subsection{Local energy density} \label{locale}

The state $\omega_a$ (unlike the state $\omega_a^B$) is defined
on the whole algebra. In the language used in \cite{her05} the
energy density in this state (properly normally ordered with
respect to the vacuum) is given in the whole space by
point-splitting procedure by
\begin{equation}
 \mc{E}_a(\vec{x})=T_a(\vec{x},\vec{x})\,,
\end{equation}
where $T_a(\vec{x},\vec{y})$ is the distribution defined by
\begin{equation}
 T_a(\vp,\psi)=\tfrac{1}{4}\big(\vp,(h_a-h)\psi\big)
 +\tfrac{1}{4}\big(\vec{\nabla}\vp,(h_a^{-1}-h^{-1})\vec{\nabla}\psi\big)\,,
\end{equation}
with scalar product between the two gradients understood in the
second term. The test functions are taken to be real. If one
takes into account the translational symmetry in the directions
parallel to the plates one realizes that the
$\vec{x}_\perp,\vec{y}_\perp$-dependence of $T_a$ may be only
through the difference $\vec{x}_\perp-\vec{y}_\perp$. The
removal of point-splitting in these directions means putting
this variable equal to zero, or integrating the $2$-dimensional
Fourier transform of $T_a$ over all space of
$\vec{p}_\perp$-variables of the spectral representation of
$h_\perp$. In this way one finds (now $x,y$ are variables in
the direction orthogonal to plates and $\vp,\psi$ are
one-dimensional)
\begin{equation}
 \mc{E}_a(x)=T_{za}(x,x)\,,
\end{equation}
where
\begin{multline*}
 T_{za}(\vp,\psi)=\frac{1}{16\pi^2}\int\bigg\{
 \Big(\vp,\big[(h_{za}^2+\vec{p}_\perp^{\,\,2})^{1/2}
 -(h_{z}^2+\vec{p}_\perp^{\,\,2})^{1/2}\big]\psi\Big)\\
 +\vec{p}_\perp^{\,\,2}
 \Big(\vp,\big[(h_{za}^2+\vec{p}_\perp^{\,\,2})^{-1/2}
 -(h_{z}^2+\vec{p}_\perp^{\,\,2})^{-1/2}\big]\psi\Big)\\
 +\Big(\vp',\big[(h_{za}^2+\vec{p}_\perp^{\,\,2})^{-1/2}
 -(h_{z}^2+\vec{p}_\perp^{\,\,2})^{-1/2}\big]\psi'\Big)\bigg\}
 d^2p_\perp\,,
\end{multline*}
with prime in $\vp',\psi'$ denoting the derivative. The
integral is easily carried out explicitly and one finds
\begin{equation}
 T_{za}(\vp,\psi)=\frac{1}{24\pi}(\vp,(h_{za}^3-h_z^3)\psi)
 -\frac{1}{8\pi}(\vp',(h_{za}-h_z)\psi')\,.
\end{equation}

Our objective is to find the limit of the energy density in the
scaled models. Thus following the introductory remarks of the
present section we assume now the supports of $\vp,\psi$ to be
outside the set $\<-b-R,-b+R\>\cup\<b-R,b+R\>$. For such
functions there is $h_{za}^2\psi=h_z^2\psi$, so
\begin{equation}\label{Tzaout}
 T_{za}(\vp,\psi)=-\frac{1}{24\pi}(\vp,(h_{za}-h_z)\psi'')
 -\frac{1}{8\pi}(\vp',(h_{za}-h_z)\psi')\,.
\end{equation}
Consider the general element $(\vp,(h_{za}-h_z)\psi)$ with test
functions in the assumed class. For any non-negative real
numbers $a,b$ one has the identity
\begin{equation*}
 a-b=-\frac{2}{\pi}\int_0^\infty\bigg\{\frac{1}{(a^2+r^2)}
 -\frac{1}{(b^2+r^2)}\bigg\}\,r^2\,dr\,.
\end{equation*}
Using this and the spectral representations of $h_{za}$ and
$h_z$ one finds
\begin{equation*}
 \big(\vp,(h_{za}-h_z)\psi\big)
 =\frac{2}{\pi}\int_0^\infty\big(\vp,\big[G(-r^2)-G_0(-r^2)\big]\psi\big)
 r^2dr\,.
\end{equation*}
To find the integral kernel $[G(-r^2)-G_0(-r^2)](x,y)$ of
$[G(-r^2)-G_0(-r^2)]$ one needs only to transform the kernel
\eqref{Gmom} to the position space:
\begin{equation*}
 [G(-r^2)-G_0(-r^2)](x,y)\\
 =\frac{1}{2\pi}\int e^{ipx} \frac{\mc{F}_p}{(p^2+r^2)}dp\
 \mc{T}(-r^2)\int e^{-iqy} \frac{\mc{F}_q^{\dagger}}{(q^2+r^2)}dq\,.
\end{equation*}
For $x,y$ in the assumed region the Fourier integrals may be
closed to contour integrals in the complex plane (the
half-circular contributions vanish either in the upper or the
lower half-plane due to the bound \eqref{csmom}) and evaluated
by residues. Because eventually we are interested in removal of
point-splitting we assume that $x$ and $y$ are in the same
connected part of the region considered. We denote
 $\mc{T}(-r^2) =
 \begin{pmatrix}
 A&B\\
 B&A
 \end{pmatrix}
 $,
where $A$ and $B$ are functions of $r$, and get
\begin{multline*}
 \big[G(-r^2)-G_0(-r^2)\big](x,y)
 =\frac{\pi}{r^{1+\si}}|\wh{f}(ir)|^2\times \\
 \times
 \begin{cases}
 e^{-|x+y|r}[A\cosh(ar)+B] &\text{for }x,y>b+R\\[-.5ex]
 &\text{or }\,\,x,y<-b-R\,,\\
 e^{-ar}\big[A\cosh\big((x+y)r\big)+\si B\cosh\big((x-y)r\big)\big]
 & \text{for }x,y\in(-b+R,b-R)\,.
 \end{cases}
\end{multline*}
We use this integral kernel in \eqref{Tzaout}, evaluate
derivatives by parts and remove the point splitting. In this
way we find in the assumed regions
\begin{multline*}
 \mc{E}_a(x)=T_{za}(x,x)\\
 =-\frac{1}{6\pi}\int_0^\infty
 \left\{
 \begin{gathered}
 2e^{-2|x|r}\big(A\cosh(ar)+B\big)\\
 e^{-ar}\big(2A\cosh(2xr)-\si B\big)
 \end{gathered}
 \right\}|\wh{f}(ir)|^2\,r^{3-\si}dr\,,\qquad
 \begin{aligned}&\text{for } |x|>b+R\,,\\
 &\text{for } |x|<b-R\,.
 \end{aligned}
\end{multline*}
We now want to consider the limit of this local energy in the
scaled version of the model. Because of  the appropriate
convergence of the integral this limit may be performed inside
the integral. The scaling of $\wh{f}(ir)$ and $\mc{T}(-r^2)$
follows from Section~\ref{odtw}. As for the scaled model the
excluded position set shrinks to $||x|-b|\leq\la R$, in the
limit all $x\neq\pm b$ are admitted. A straightforward
calculation yields
\begin{equation*}
 \liml \mc{E}_{a,\la}(x)
 =-\frac{\si}{6\pi^2}\int_0^\infty
 \Bigg\{
 \begin{gathered}
 e^{-(2|x|-a)r}\\
 \frac{e^{(2x-a)r}+e^{-(2x+a)r}+\si e^{-2ar}}{1-e^{-2ar}}
 \end{gathered}
 \Bigg\}\,
 r^3 dr\,,\qquad
 \begin{aligned}&\text{for } |x|>b\,,\\[1.5ex]
 &\text{for } |x|<b\,.
 \end{aligned}
\end{equation*}
With the help of Eq.\ \eqref{Hzeta} in Appendix \ref{P} we get
\begin{equation}\label{ElimB}
 \mc{E}_a^B(x)\equiv
 \liml \mc{E}_{a,\la}(x)\\
 =
 \begin{cases}
 -\,\dfrac{\si}{16\pi^2(|x|-b)^4}&\text{for }|x|>b\,,\\
 -\,\dfrac{\pi^2}{1440a^4}\,
 -\displaystyle\sum_{n\in(2\mbb{Z}+1)}
 \dfrac{\si}{16\pi^2(nb-x)^4}
 &\text{for }|x|<b\,,
 \end{cases}
\end{equation}
where $2\mbb{Z}+1$ denotes the set of odd whole numbers.

Let us stress once more: the above result holds in the
distributional sense only for test functions supported outside
$x=\pm b$, i.e.\ for functions in that class there is
\begin{equation*}
 \liml \int\mc{E}_{a,\la}(x)\vp(x)dx
 =\int\mc{E}_a^B(x)\vp(x)dx\,.
\end{equation*}
For functions not in this class our assumptions leading to the
above result cease to hold. Nevertheless, we shall attempt some
comparison with our global results. For that purpose let us
denote
\begin{equation}\label{EBone}
 \mc{E}^B(x)=-\frac{\si}{16\pi^2x^4}\qquad
 \text{for }x\neq0\,.
\end{equation}
Then \eqref{ElimB} may be written as
\begin{equation}\label{EBint}
 \mc{E}_a^B(x)=\mc{E}^B(x+b)+
 \mc{E}^B(x-b)+\mc{E}_{a,\mr{int}}^B(x)\,,
\end{equation}
where
\begin{equation}
 \mc{E}_{a,\mr{int}}^B(x) =
 \begin{cases}
 +\,\dfrac{\si}{16\pi^2(|x|+b)^4}&\text{for }|x|>b\,,\\
 -\,\dfrac{\pi^2}{1440a^4}\,
 -\displaystyle\sum_{n\in(2\mbb{Z}+1)\setminus\{+1,-1\}}
 \dfrac{\si}{16\pi^2(nb-x)^4}
 &\text{for }|x|<b\,.
 \end{cases}
\end{equation}
It is easy to see that $\mc{E}_{a,\mr{int}}^B(x)=a^{-4}F(x/a)$,
where $F(y)$ is an absolutely bounded and integrable function.
Thus for large separation of plates the energy outside the
plates is concentrated in the first two terms in \eqref{EBint}.
Therefore \eqref{EBone} has the interpretation of the energy
density produced around one single plate, while
$\mc{E}_{a,\mr{int}}^B(x)$ may be regarded as the energy
density (\emph{locally outside the boundaries}) of the
interaction. When integrated over $x\in\mbb{R}$ it gives
\begin{equation}\label{Eintint}
 \mc{E}_{a,\mr{int}}^B=-\frac{\pi^2}{1440 a^3}
\end{equation}
for both Dirichlet and Neumann cases.

\section{Discussion}\label{discussion}

The models considered in this article do not pretend to
describe the details of the interaction of quantum field with
macroscopic bodies in realistic way. Their merit comes from the
fact that (i) they are consistent with the restrictions imposed
by the general algebraic analysis of any quantum backreaction
setting \cite{her05} while (ii) being simple enough to allow
explicit calculations, and (iii) approximating sharp boundary
conditions in a controllably localized (in physical space) way.
The last point distinguishes them from the class of models
constructed in \cite{her06} and allows the comparison of the
results of the global and local analyzes.

We summarize the results and lessons to be drawn from them.

\vspace{1ex} \noindent
 (i) The models discussed in the present article are
defined on the Weyl algebra of the free theory, and the
interaction introduced by the `nonlocal boundaries' does not
lead out of the vacuum representation of the theory. The energy
of the field (as defined by the free theory) is finite in the
ground state (as defined by the interaction with `boundaries')
without any arbitrary `renormalization'. The change of this
`Casimir energy' with the variation of the position of the
boundaries determines the backreaction force.

\vspace{1ex} \noindent
 (ii) The scaled interaction with the `nonlocal boundaries' approaches
 in the scaling limit the sharp Dirichlet/Neumann conditions in
 the Hilbert-Schmidt norm sense for the resolvents. This
 implies norm convergence (or strong convergence) for any continuous,
 vanishing in infinity (resp.\ any bounded) function of the
 first-quantized hamiltonian. The nonlocality of the boundaries
 is under control and tends to zero in the limit.

\vspace{1ex} \noindent
 (iii) The Casimir energy per area of the scaled models obeys the
 scaling law \eqref{en-scale}. Thus it is governed by the
 $a$-dependence of the energy given in Eqs.\ \eqref{CED} and
 \eqref{CEN}. The Casimir force per area is minus the derivative of those
 formulas. We have not discussed this point, but one can show
 that one can differentiate terms in these formulas one by one
 with $o(a^{-3})$ going over to $o(a^{-4})$. Thus one finds
\begin{equation*}
 -\frac{d\ve_a^\la}{da}=
 \begin{cases}
 -\dfrac{\pi^2}{480 a^4}+o\Big(\dfrac{\la}{a^{4}}\Big)\,,& \text{(D)}\\
 +\dfrac{1}{24\pi M_0\la a^3}
 +\dfrac{I_0}{8\pi^2 M_0^2 a^4}\left(\dfrac{3\zeta(3)}{\pi^2}+1\right)
 -\dfrac{\pi^2}{480 a^4}+ o\Big(\dfrac{\la}{a^{4}}\Big)\,.&\text{(N)}
 \end{cases}
\end{equation*}
One finds that in Dirichlet case the force has a well-defined
limit, but in Neumann case depends on the model and diverges
for sharp boundaries. This model-dependence occurs despite the
fact that the models approximate in many respects the sharp
boundaries very well (globally!). Neumann case models with
$I_0=0$ (which are among those admitted by our assumptions --
see Appendix C) have faster convergence property (see Eq.\
\eqref{weakN}) and for them the additional $a^{-4}$ term in the
force vanishes.

\vspace{1ex} \noindent
 (iv) Algebras of fields localized outside the interaction
 region (test functions with supports not intersecting with
 that region) admit free vacuum state, ground states of our models,
 as well as ground state of the sharp boundaries. Restricted
 to these algebras our ground states tend weakly to the sharp
 boundaries states in the scaling limit.

\vspace{1ex} \noindent
 (v) The local energy density is unambiguously defined in our
 models in the whole physical space by point splitting, with no
 ad hoc later renormalization. When restricted to regions not
 intersecting with the interaction area (which in scaling limit
 just means not intersecting with the boundaries) the local
 Casimir energy has a well defined limit given by a smooth
 function Eq.\ \eqref{ElimB}. This limit density has universal,
 model-independent form. Thus the `bulk' contribution to the
 total Casimir energy has this universality and the
 model-dependent terms in Neumann case turn out to be squeezed
 in the limit inside the boundaries. The hope that these
 non-regular contributions may be removed in a
 model-independent way is therefore not justified. Further
 confirmation of our interpretation supply formulas and remarks
 ending the last section.

 Local energy density has been discussed by various authors
 before \cite{local}--\cite{kaw07}, but usually with the use of some regularizations,
 often of not quite clear status. The results to be found in
 literature are not consistent. The Dirichlet case for the
 region between the plates is discussed in \cite{mil03} and
 modulo some infinite renormalization agrees with ours
 (formula (2.32) in that reference). On the
 other hand the authors of \cite{kaw07} obtain a different
 result (by a rather indirect way of `regularization' and
 removal of cut-off). We are not aware of a complete rigorous
 discussion resulting in our
 formulas~\eqref{ElimB}~--~\eqref{Eintint}. It is also worth
 noting that the density $\mc{E}_a^B(x)$ can be also obtained
 directly by the use of sharp boundary conditions hamiltonian
 $h_{za}^B$ instead of $h_{za}$ in \eqref{Tzaout}. This amounts
 to the use of the difference of \eqref{GB} and \eqref{G0} in
 the calculation. However, let us stress once more, this limit
 value of the density is correct only if smeared with a test
 function with support not touching the borders.\\

\noindent
 We are grateful to the referee for careful reading
and constructive editorial remarks.

\section*{Appendices}
\appendix

\appendix

If not stated otherwise we work here with the same assumptions
as stated in \eqref{assum} and \eqref{normN}.

\section{Integrals and estimates}
\label{P}

We gather here a few technical, separate points used in the
main text.\\

\noindent {\bf(i)}\quad
 We recall that $f$ is a smooth function supported in
$\<-R,R\>$ if, and only if, its Fourier transform $\wh{f}$ is
an entire analytic function satisfying the estimates
\begin{equation}\label{csmom}
 |\wh{f}(u)|\le\frac{\con(N)\,e^{R|\Imaginalis u|}}{(1+|u|)^N} \,,
\end{equation}
for all $u\in\mbb{C}$ and $N\in\mbb{N}$.

Therefore $M_p$, which is the product of the Fourier transforms
of $f(z)$ and of $\ol{f(-z)}=\ol{f(z)}$, extends to the
analytic function $M_u$ on $\mbb{C}$ satisfying similar
estimates with $R$ replaced by $2R$. The inverse transform
$\check{M}$ of $M_p$ is smooth and supported in $\<-2R,2R\>$.
Using this we find for $a>2R$
\begin{equation} \label{cosM}
 \int \cos(ap)M_pdp=\sqrt{\frac{\pi}{2}}
 \left[\check{M}(a)+\check{M}(-a)\right]=0\,.
\end{equation}
As $M_u$ is even, $\displaystyle\frac{M_p-M_k}{p^2-k^2}$ also
extends to an analytic function and with the use of estimates
on $M_u$ one finds by closing the contour of integration (as
always for $a>2R$) that
\begin{equation} \label{cosIk}
 \int \frac{\cos(ap)(M_p-M_k)}{p^2-k^2}dp=0\,.
\end{equation}\\

\noindent {\bf(ii)}\quad In order to prove the estimate
\eqref{D3/2}, we start with Dirichlet case, we note that
(remember that $\wh{f}'(0)=0$, as $\wh{f}$ is even)
\begin{equation*}
 \bigg|\frac{\wh{f}(\la p) - \wh{f}(0)}{p^2}\bigg|
 =\Bigg| \int_0^\la \wh{f}''(p\xi)(\la-\xi)d\xi \Bigg|
 \le \la \int_0^\la \big|\wh{f}''(p\xi)\big|d\xi\,.
\end{equation*}
Moreover we have (norms of functions of $p$ as elements of
$L^2$)
\begin{multline*}
 \Bigg\|\int_0^\la \big|\wh{f}''(p\xi)\big|d\xi \Bigg\|^2
 \le \int_{\<0,\la\>^2} \big\|\wh{f}''(p\xi)\big\| \big\|\wh{f}''(p\tilde{\xi})\big\|
 d\xi d\tilde{\xi}\\
 \le \big\|\wh{f}''\big\|^2 \int_{\<0,\la\>^2} (\xi \tilde{\xi})^{-1/2}
 d\xi d\tilde{\xi} \le \con\, \la \,,
\end{multline*}
where in the first step we have used the Schwartz inequality.
Now because $\displaystyle\frac{p^2}{w^2-p^2}$ is bounded we
get the estimate for Dirichlet. The same considerations show
\eqref{N3/2}. The proof of Neumann case in \eqref{D3/2} is
almost the same but with the use of the formula
\begin{equation*}
 \wh{f}(\la p) - \wh{f}(0) = p \int_0^\la \wh{f}'(p\xi)d\xi\,.
\end{equation*}\\

\noindent {\bf(iii)}\quad For the function $I_k$, \eqref{Ik},
we first observe that for some $0<\ep_0<1$ there is
\begin{equation} \label{estI0}
 I_0<2\pi RM_0(1-\ep_0)\,.
\end{equation}
This follows from the identity
\begin{equation*}
 4\int_0^\infty\frac{\sin^2(Rp)M_p}{p^2}dp=2\pi RM_0-I_0\,,
\end{equation*}
(used already in the normalization of the bound state, see
\eqref{psi0},\eqref{psi0N}), as the l.h.s. is strictly greater
than zero. Further, we need the following estimates:
\begin{equation} \label{estIk}
 \begin{aligned}
 &I_k\geq\frac{\con(k_*)}{k^2}\,,& &k\ge k_*\,,\\
 &I_k\le\frac{\con}{(1+k)^2}\,,& &k\geq0 \,,
 \end{aligned}
\end{equation}
with arbitrary $k_*>0$. We write $I_k$ as a principal value
distribution calculated on test function~$M$. In position
space, using evenness of $f$, we have
\begin{equation} \label{Ikinx}
 I_k=\sqrt{2\pi}\int_0^\infty x\check{M}(x)\frac{\sin(kx)}{kx}dx\,.
\end{equation}
Integrating once by parts we get
\begin{equation} \label{estIk2}
 I_k=\frac{\sqrt{2\pi}}{k^2}\bigg[\check{M}(0)+\int_0^\infty
 \check{M}'(x)\cos(kx)dx\bigg]\,.
\end{equation}
The estimation from above is now trivial, whereas for the
estimation from below we use the Riemann-Lebesgue lemma, the
assumption that $I_k>0$ for $k\neq0$ (see \eqref{Irange}) and
continuity of $I_k$. Expanding further in powers of $1/k$ the
integral in \eqref{estIk2} we find moreover
\begin{equation}\label{estdI}
 |\partial_k^n I_k|\leq \frac{\con}{(1+k)^{n+2}}\,.
\end{equation}

 \noindent {\bf(iv)}\quad  The $s_k$ function, defined
in the end of Section \ref{spectral}, is smooth for $k\ge0$ and
satisfies
\begin{equation} \label{s_k}
|\partial_k^n s_k|\le \con(n)\, (1+k)^{-(n+\si)}\,.
\end{equation}
To show this we note first that $|M_k-iN_k|^{-1}$ is bounded in
a neighborhood of $k=0$ as $M_0>0$. Outside this neighborhood
we have $|M_k-iN_k|^{-1}\leq|N_k|^{-1}\le \con \,(1+k)^{-\si}$
due to \eqref{IrangeD} and the first bound in \eqref{estIk}. On
the other hand due to \eqref{estdI} there is $|\partial_k^n
N_k|\leq\con(1+k)^{-(n+1)}$ for $n\geq1$ except for $n=1$ in
the Dirichlet case, when this is replaced by
$|\partial_kN_k|\leq\con$. As $M_k$ is Schwartz, this ends the
proof.

It now follows that $q_k=s_kM_k$ and all its derivatives are
bounded by Schwartz functions.

\noindent {\bf(v)}\quad  For both (D and N) cases and $k\ge0$
we note the bound
\begin{equation} \label{est_denom}
 \frac{1}{\big|1-\left(q_k e^{iak}\right)^2\big|}
 \le\con(a)\,\frac{1+k}{k}\,.
\end{equation}
At $k=0$ there is  $1-(q_0)^2=0$ but the first derivative of
the denominator at zero does not vanish (in Neumann case use
\eqref{estI0}), which is sufficient for \eqref{est_denom} in a
neighborhood of zero. Outside that neighborhood, mainly due to
\eqref{estIk}, it follows that
\begin{equation*}
 \frac{1}{\big|1-\left(q_k e^{iak}\right)^2\big|}
 \le\frac{1}{1-|q_k|^2}=\frac{M_k^2+N_k^2}{N_k^2}
 \le\con\,,
\end{equation*}
which ends the proof. We also note that for $k\ge0$ we have
\begin{equation} \label{estden2}
 \frac{1}{1-|q_k|^2}\le\con\,\frac{1+k^m}{k^m}\,,
 \quad
 \begin{cases}
 m=2\,,&\text{(D)}\\
 m\geq2\,,&\text{(N)}
 \end{cases}
\end{equation}
where the Neumann case depends on the behavior of $I_k$ at zero
($m=2$ for $I_0\neq0$ and $m=2+2r$ when $I_k\underset{k \rightarrow
0}{\simeq}k^{2r}$, $r\ge1$ as $I_k$ is even).\\

\noindent {\bf(vi)}\quad  Finally, we note the following
identity using a known integral representation of the Hurwitz
zeta function: for $\alpha>0$ there is
\begin{equation} \label{Hzeta}
 \frac{1}{6}\int_0^\infty \frac{r^3 e^{-\alpha r}}{1-e^{-2ar}} dr
 =\frac{1}{(2a)^4} \zeta\Big(4,\frac{\alpha}{2a}\Big)
 =\sum_{n=0}^\infty (\alpha+2an)^{-4}\,,
\end{equation}
which is needed for the calculation of \eqref{ElimB}.

\section{Operations (i) -- (iii) from Section \ref{enexp}}
\label{chi'''}

In this appendix we prove the admissibility of the three
operations (i) -- (iii) performed in Section \ref{enexpD} for
the Dirichlet case and mentioned in Section \ref{enexpN} for
the Neumann case. The key tool for this is the following simple
lemma:

Let $c_n$ for $n\in\mathscr{N}\subseteq\mbb{N}$ be complex
measurable functions on $D\subseteq\mbb{R}$.
 If~$\sum_{n\in\mathscr{N}}|c_n (k)|$ is integrable on $D$ then
\begin{equation} \label{lemat}
 \lim_{a \rightarrow \infty} \sum_{n \in \mathscr{N}} \int_D c_n (k)
 e^{inak} dk = 0 \,.
\end{equation}
The proof is straightforward and uses the Lebesgue dominated
convergence theorem and the Riemann-Lebesgue lemma. Throughout
this appendix $\mathscr{S}_k$ denotes some Schwartz function in $k$
variable (each time it may be a different function). All we
have to do is to check if the assumption of the mentioned lemma
is fulfilled for the appropriate expressions.

\subsection{Dirichlet case}

Let us first consider the following part of the expression
\eqref{energyD}
\begin{equation} \label{AiD}
 \frac{1}{a^3}\sum_{n\in2\mbb{N}}\frac{i}{n^3}\int_{\mbb{R}_+^2}
 M_p\partial_k^3 \Bigl(\chi(k,p)s_k\,q_k^n \Bigr)e^{inak}dk\,dp\,;
\end{equation}
we shall need its real part multiplied by $-1/6\pi^3$ in
\eqref{energyD}.

\noindent {\bf(i)}\quad To justify the validity of operation
(i) of Section \ref{enexpD} it is sufficient to check whether
the sequence of functions
\begin{equation*}
 c_n(k)=n^{-3}\int_0^\infty M_p
 \Big[\partial_k^3 \bigl(s_kq_k^n\,\chi(k,p) \bigr)
 -s_kq_k^n\,\partial_k^3 \chi(k,p)\Big]\,dp
\end{equation*}
satisfies the assumption of lemma \eqref{lemat}.  For this we
note that using properties of $s_k$ and $q_k$ described in
Appendix A (iv) we have
\begin{multline}\label{nnn}
 n^{-3}\Big|\partial_k^3 \bigl(s_kq_k^n\,\chi(k,p) \bigr)
 -s_kq_k^n\,\partial_k^3 \chi(k,p)\Big|\\
 \leq
 \mathscr{S}_k\Big(n^{-3}|q_k|^{n-1}+n^{-2}|q_k|^{n-1}\Big)
 \sum_{j=0}^2|\partial_k^j\chi(k,p)|\\
 +\mathscr{S}_kn^{-2}(n-1)|q_k|^{n-2}\sum_{j=0}^1|\partial_k^j\chi(k,p)|
 +\mathscr{S}_kn^{-2}(n-1)(n-2)|q_k|^{n-3}|\chi(k,p)|\,.
\end{multline}
A simple calculation yields
\begin{equation}\label{estchi}
 |\partial_k^j\chi(k,p)|\leq\con\, \frac{k^{2-j}}{k+p}\,,
 \qquad j=0,1,2\,,
\end{equation}
and therefore
\begin{equation}\label{intchi}
 \int_0^\infty M_p|\partial_k^j\chi(k,p)|dp\leq\con\,
 k^{2-j}\log(1+k^{-1})\,,\qquad j=0,1,2\,.
\end{equation}
Using this and remembering that $|q_k|\leq1$ we obtain
\begin{equation}\label{modc}
 |c_n(k)|\leq \mathscr{S}_k\log\big(1+k^{-1}\big)
 \Big(n^{-3}+n^{-2}+ n^{-1}k|q_k|^{n-2}+k^2|q_k|^{n-3}\Big)\,,
\end{equation}
where the third and the fourth terms inside the parentheses
appear for $n\geq2$ and $n\geq3$ respectively. The summation
gives
\begin{equation}\label{summodc}
 \sum_{n\in2\mbb{N}}|c_n(k)|\leq \mathscr{S}_k\log\big(1+k^{-1}\big)
 \bigg[1+k\log\bigg(\frac{1}{1-|q_k|^2}\bigg)+\frac{k^2}{1-|q_k|^2}\bigg]\,.
\end{equation}
The estimate \eqref{estden2} shows that the integral of the
r.h.s. over $\mbb{R}_+$ is finite, thus the assumption of lemma
\eqref{lemat} is satisfied.

\noindent {\bf(ii)}\quad For the operation (ii) the assumption
of the lemma is fulfilled because
\begin{equation*}
 \left| M_p s_k q_k^n \partial_k^3\chi(k,p) \right|
 \le \frac{6p^2(k+3p)}{(k+p)^5} M_p \mathscr{S}_k
 \in L^1\big(\mbb{R}_+^2 \backslash \, \Omega\big) \,,
\end{equation*}
with $\Omega$ as defined in Section \ref{enexpD}.

\noindent {\bf(iii)}\quad The operation (iii) is admissible
because
\begin{equation*}
 |M_p s_k q_k^n - M_0 s_0 q_0^n|
 \le \con \: p + \con\, (1+n) k + \con\, (1+n)pk
\end{equation*}
and $k\,\partial_k^3\chi(k,p), p\,\partial_k^3\chi(k,p)$ are both integrable on
$\Omega$, so the lemma holds also in this case. This ends the
proof for the terms \eqref{AiD}.

For the expression with the sum over odd natural numbers and
with $\cos(ap)$ we use the following modification of the lemma.
Let $c_n$ for $n\in\mathscr{N}\subseteq\mbb{N}$ be complex
measurable functions on $D\subseteq\mbb{R}^2$.
If~$\sum_{n\in\mathscr{N}}|c_n (k,p)|$ is integrable on $D$
then
\begin{equation} \label{lemat2}
 \lim_{a \rightarrow \infty} \sum_{n \in \mathscr{N}} \int_D c_n (k,p)
 e^{inak} e^{\pm iap} dk dp = 0 \,.
\end{equation}
Now, almost the same considerations as before show the
admissibility of operations (i) -- (iii) for this part of
energy.

\subsection{Neumann case}
First we note that for the terms which we consider here, the
limit over $\epsilon$ can be easily performed. This follows
from the estimations below. Therefore we use the same lemma as
for Dirichlet case, i.e. \eqref{lemat} and \eqref{lemat2}, but
we recall that we now replace $q_k e^{iak} = \tilde{q}_k
e^{i\tilde{a}k}$, as mentioned at the beginning of Section
\ref{enexpN}. With this modification the estimation \eqref{nnn}
for the terms in the sum \eqref{AiD} is still valid, but the
bounds \eqref{estchi} change: $k^{2-j}$ is replaced by
$p^{2-j}$. In consequence there is no $k^{2-j}$ factor in front
of $\log$ in \eqref{intchi} and no $k$, $k^2$ factors in
\eqref{modc} and \eqref{summodc}. With this modification the
sum of $n^{-3}$, $n^{-2}$ and $n^{-1}$ terms in \eqref{modc} is
still sufficiently well bounded, but the sum of $n^0$ terms is
to singular (no $k^2$ in the last term in \eqref{summodc}).
Therefore these terms need a more detailed treatment. In fact,
they can be estimated by \mbox{$\con\,|s_k| |\tilde{q}_k|^{n-3}
|\tilde{q}_k'|^3\log(1+k^{-1})$} (prime denotes here the
derivative with respect to $k$) and their sum over $n$ is
bounded by
\begin{equation*}
 \con\,\frac{|s_k||\tilde{q}_k'|^3}{1-|\tilde{q}_k|^2}\log(1+k^{-1})\,.
\end{equation*}
This function is indeed in $L^1(\mbb{R}_+)$, since outside the
neighborhood of zero, using \eqref{estden2}, we have
$\displaystyle\frac{|s_k||\tilde{q}_k'|^3}{1-|\tilde{q}_k|^2}\le\mathscr{S}_k$,
whereas in the neighborhood we have $\tilde{q}_0^\prime=0$
which is enough for $I_0\neq0$ case (see \eqref{estden2} and
the comment after it) and if $I_k\underset{k \rightarrow
0}{\simeq}k^{2r}$, $r\ge1$ ($I_k$ is even), then
$\displaystyle\frac{|\tilde{q}_k'|^3}{1-|\tilde{q}_k|^2}\underset{k
\rightarrow 0}{\simeq}k^{2r-2}$. The discussion of
admissibility of operations (ii) and (iii) goes in the same way
as for Dirichlet case (estimates hold also for Neumann case,
with $\tilde{q}_k$ instead of $q_k$). The analysis of the
expression with the sum over odd natural numbers and with
$\cos(ap)$ is also analogous to the Dirichlet case.

\section{Comments on the assumed class of models}

The models discussed in this paper are based on functions $f$
subject to the conditions formulated in \eqref{assum},
\eqref{normN}, \eqref{Irange} and \eqref{IrangeD}. In this
appendix we exhibit a class of functions conforming to them. It
is sufficient to satisfy \eqref{assum} and \eqref{Irange} as
the other two are then achieved by simple rescaling of the
function by a constant factor.

First, we note that each even function with the assumed support
which in addition is real, non-negative and monotonically
(weakly) decreasing for positive arguments satisfies the
demands. Indeed, for each function the last condition in
\eqref{assum} is fulfilled. Moreover, it is easy to see that
then $\check{M}(x)$ (being the convolution of the function with
itself) is also even, positive, compactly supported and
decreasing for $x>0$. Thus from \eqref{estIk2}, since for
$k\neq0$ there is
\begin{equation*}
 \bigg|\int_0^\infty\check{M}'(x)\cos(kx)dx\bigg|
 <-\int_0^\infty\check{M}'(x)dx
 =\check{M}(0)\,,
\end{equation*}
we have $I_k>0$ for $k\neq0$, which ends the proof of
\eqref{Irange}.

For each function in the class defined in the previous
paragraph there is $I_0>0$, which is due to the positivity of
$\check{M}(x)$ (see \eqref{Ikinx}). We now extend the class to
include also functions for which $I_0=0$ (see Discussion). Let
$f$ be a function in the class of the last paragraph and define
a new function
\begin{equation*}
 f^r(z)=f(z)-\mu\big(f(z-r)+f(z+r)\big)\,,
\end{equation*}
where $\mu >0$ and $r>R>0$. One finds
\begin{equation*}
 M^r_p=|\wh{f}^r(p)|^2=\big(1-2\mu\cos(rp)\big)^2 M_p\,.
\end{equation*}
Using this and taking into account \eqref{cosIk} one has
\begin{equation*}
 I^r_k=
 \int\frac{M^r_k-M^r_p}{p^2-k^2}dp=\Big(1+2\mu^2\Big)I_k
 -4\pi\mu\frac{\sin(rk)}{k}\left(1-\mu\cos(rk)\right)M_k\,.
\end{equation*}
We impose the condition $I^r_0 = 0$, which is a quadratic
equation for $\mu$:
\begin{equation*}
 2(I_0+2\pi rM_0)\mu^2-4\pi rM_0\mu+I_0=0\,.
\end{equation*}
For sufficiently large $r$ the equation has two roots, and we
take the smaller one, which is less then $1/2$ and for large
$r$ tends to zero. Then
\begin{equation*}
 I^r_k=4\pi r\mu(1-\mu)M_0\frac{I_k}{I_0}\big[1-\eta(rk)\xi(k)\big]\,,
\end{equation*}
where
\begin{equation*}
 \eta(rk)=\frac{\sin(rk)}{rk}\,\frac{1-\mu\cos(rk)}{1-\mu}\,,\q\xi(k)
 =\frac{I_0 M_k}{M_0 I_k} \,.
\end{equation*}
It is an exercise in function analysis to show that for
sufficiently small $\mu$ there is $\eta(u)<1$ for all $u\neq0$.
Below we show that $f$ may be chosen such that also $\xi(k)<1$
for $k\neq0$, and then $I_k^r>0$ for $k>0$, which is the
condition \eqref{Irange}. To reduce the size of the support of
$f^r$ one can use the scaling defined in Section \ref{odtw}.

Consider now the non-negative and even function $\xi(k)$ for
positive arguments. Suppose that $\xi''(0)<0$ (see below). Then
there is $k_0>0$ such that for $k \in (0, k_0 )$ it is
$\xi(k)<1$. For $k\ge k_0$, using $\eqref{estIk}$, we have
$\xi(k)\le\con\equiv\xi_\mr{max}$. We now need to improve the
estimation of $\eta$ for $k\ge k_0$. It is easy to see that if
$r$ has been chosen large enough then $\eta(rk) <
\frac{1}{\xi_\mr{max}}$ for $k\ge k_0$.

Finally, we have to choose $f$ so as to satisfy $\xi''(0)<0$.
Let $f$ be defined as $f(z) = f_0=\con$\ on
 $\left\<-R,R\right\>$ and zero outside. Straightforward calculation
gives
\begin{equation*}
 M_k = \frac{2|f_0|^2}{\pi}R^2\frac{\sin^2 (Rk)}{(Rk)^2}\,,\quad
 I_k=8|f_0|^2 R^3\frac{2Rk-\sin(2Rk)}{(2Rk)^3}\,,\quad
 \xi''(0)=-\frac{8}{5}\,R^2<0\,.
\end{equation*}
The function used here needs `rounding the corners' to be in
the class of the second paragraph. But this may be made by a
small local variation, and both $M_k$ and $I_k$ and their
derivatives depend continuously on such small variations of $f$
(for $I_k$ see \eqref{Ikinx}), which is sufficient to conclude
the proof.


\begin{thebibliography}{99}

\frenchspacing

\bibitem{her05} A.\,Herdegen, {\sl Ann. H. Poincar\'e} {\bf 6},
    669-707 (2005).
\bibitem{her06} A.\,Herdegen, {\sl Ann. H. Poincar\'e} {\bf 7},
    253-301 (2006).
\bibitem{nonloc} A.\,Jaffe and L.R.\,Williamson, {\sl Annals
    Phys.} {\bf 282} 432 (2000);\\
    C.D.\,Fosco and E.\,Losada, {\sl Phys. Lett. B} {\bf 675} 252 (2009).
\bibitem{local} C.A.\,L\"utken and F.\,Ravndal, {\sl Phys.
    Rev. A} {\bf 31}, 2082 (1985);\\
    V.\,Sopova and L.H.\,Ford {\sl Phys. Rev.} {\bf D66},
    045026 (2002).
\bibitem{mil03} K.A.\,Milton, {\sl Phys. Rev.} {\bf D68},
    065020 (2003).
\bibitem{kaw07} N.A.\,Kawakami, M.C.\,Nemes and
    W.F.\,Wreszinski, {\sl J. Math. Phys.} {\bf 48}, 102302
    (2007).

\end{thebibliography}
\end{document}